\documentclass[aps,prd,floatfix,noshowpacs,twocolumn,10pt,showkeys]{revtex4-1}

 \usepackage{epsfig}
\usepackage{epstopdf}

\usepackage[usenames,dvipsnames]{color}

\definecolor{webgreen}{rgb}{0,0.75,0}
\definecolor{webred}{rgb}{0.75,0,0}
\definecolor{webblue}{rgb}{0,0,0.75}
\definecolor{darkblue}{rgb}{0,0,0.7}
\definecolor{dunkelgrau}{rgb}{0.8,0.8,0.8}
\definecolor{lgray}{rgb}{0.95,0.95,0.95}
\definecolor{lgreen}{rgb}{0.95,1.00,0.90}
\definecolor{lblue}{rgb}{0.9,0.95,1.00}
\definecolor{lred}{rgb}{1.00,0.90,0.80}
\definecolor{shadecolor}{rgb}{1.00,0.92,0.82}

\usepackage{scalerel}

\newcommand\reallywidehat[1]{%
\begin{array}{c}
\stretchto{
  \scaleto{
    \scalerel*[\widthof{#1}]{\bigwedge}
    {\rule[-\textheight/2]{1ex}{\textheight}} %WIDTH-LIMITED BIG WEDGE
  }{1.25\textheight} % THIS STRETCHES THE WEDGE A LITTLE EXTRA WIDE
}{0.5ex}\\           % THIS SQUEEZES THE WEDGE TO 0.5ex HEIGHT
#1\\                 % THIS STACKS THE WEDGE ATOP THE ARGUMENT
\rule{-1ex}{0ex}
\end{array}
}
\usepackage{graphicx}

\usepackage{natbib}
\usepackage{multirow}
\usepackage{bbm}
\usepackage{amssymb}
\usepackage{amsmath}

\usepackage{amsfonts}
\usepackage{mathrsfs}
\usepackage{bm}
\usepackage{color}

\usepackage[colorlinks=true,linkcolor=darkblue,citecolor=darkblue,urlcolor=darkblue,breaklinks=true]{hyperref}
\def\be{\begin{equation}}
\def\ee{\end{equation}}
\def\bea{\begin{eqnarray}}
\def\eea{\end{eqnarray}}
\def\bfl{\begin{flushleft}}
\def\efl{\end{flushleft}}
\def\bfr{\begin{flushright}}
\def\efr{\end{flushright}}
\def\bc{\begin{center}}
\def\ec{\end{center}}
\def\ben{\begin{enumerate}}
\def\een{\end{enumerate}}
\def\bit{\begin{itemize}}
\def\eit{\end{itemize}}

\def\dzn{,\kern-0.1em,}

\def\Lan{\langle}
\def\Ran{\rangle}

\def\i{\text{i}}
\def\e{\mbox{e}}
\def\d{\mbox{d}}
\def\L{{\mathcal{L}}}
\def\H{{\mathcal{H}}}
\def\O{{\mathcal{O}}}
\def\I{{\mathcal{I}}}
\def\TI{\widetilde{\mathcal{I}}}

\def\Lan{\langle}
\def\Ran{\rangle}

\begin{document}

\title{Magnon-Magnon Interactions in  O(3) Ferromagnets and
Equations of Motion for Spin Operators}

\author{Slobodan M. Rado\v sevi\' c }
\email{slobodan@df.uns.ac.rs}

\affiliation{Department of Physics, Faculty of Sciences, University of Novi Sad, Trg Dositeja
 Obradovi\' ca 4, Novi Sad, Serbia}

\begin{abstract}

The method of equations of motion for spin operators in the case of O(3) Heisenberg
ferromagnet is systematically analyzed starting from the effective Lagrangian.
It is shown that the random phase approximation and the Callen approximation can be understood
in terms of perturbation theory for type B magnons. Also, the second order approximation
of Kondo and Yamaji for one dimensional ferromagnet is reduced to the perturbation
theory for type A magnons. An emphasis is put on the
physical picture, i.e. on magnon-magnon interactions and symmetries of the Heisenberg model.
Calculations demonstrate that all three approximations differ in manner in which the magnon-magnon interactions
arising from the Wess-Zumino term are treated, from where specific features  and limitations
of each of them  can be deduced.
\end{abstract}


\maketitle

%===============================================================================================================================================
\section{Introduction}
%===============================================================================================================================================

\label{intro}

The study of magnon-magnon interactions   is a problem with long 
history \cite{Zitomirski} whose complexity initiated development
of numerous analytical tools. 
Its importance goes 
far beyond pure academic interest since with the advent of new 
materials, especially those connected to the high-temperature
superconductors \cite{Dagotto,MetalInsul,PnictRMP,PnictAdvPhys},
the question of magnon-magnon interactions and their 
influence on thermodynamic properties of ordered magnets
became an urgent one to solve. 
This amounts not only to finding efficient calculation techniques,
but also  understanding how or why certain approximations work
and what are their domains  of  applicability.
With this in mind, it is useful to study simpler models \cite{WIQFT}.
In this case, they reveal some features generic to the wide class of spin systems
otherwise concealed in more complicated ones containing 
second and third neighbor interaction, various types of frustration
and anisotropies, random impurities  etc.
Much like 2D Ising model, due to several well established and
exact results, the (quantum) O(3) Heisenberg ferromagnet
has come to be a prototypical model for testing diverse
theoretical and numerical methods.

\begin{figure*}
\bc 
\includegraphics[scale=1.0]{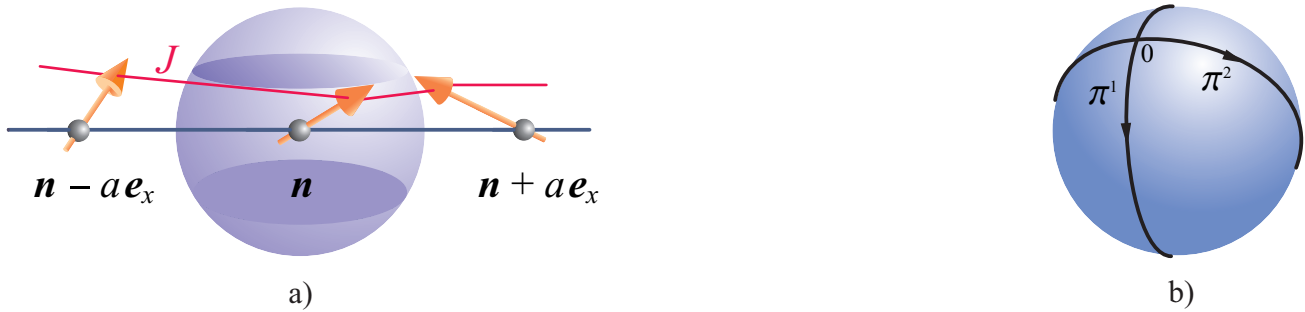}
\caption
{\label{fig1} Different perspectives on  dynamical degrees of freedom
and magnon-magnon interactions in O(3) ferromagnets: \newline
a) In spin-operator-based
approaches, interactions are viewed as a consequence of nontrivial
algebra and Hilbert space of spin operators coupled through exchange
integral $J$
[One-dimensional lattice of localized $S=1/2$ spins is displayed for clarity]\newline 
b) In EFT, magnons are considered as a set of (local) coordinates on the sphere $S^2$ = O(3)/O(2)
and magnon-magnon interactions are dictated by the underlying geometry of $S^2$, the non-Abelian
character of O(3) and the existence of conserved charges in ferromagnetic ground state.}
\ec
\end{figure*}

Nearly all theories addressing the problem of magnon-magnon interactions
are spin-operator oriented in the sense that all calculations are based
on the operators used to define  the Heisenberg model (See Fig. \ref{fig1}).
The most straightforward way to incorporate the 
interaction effects   is to express the localized spins 
in terms of bosonic/fermionic 
operators directly  in the Heisenberg Hamiltonian
or to use the coherent-state path integral
(See e.g. \cite{Auerbach,Irkin}). The original spin 
Hamiltonian is afterwards interpreted as describing 
the system of interacting bosons/fermions,
to which standard perturbation theory, or suitable
mean-field  approximations (MFA), may be applied 
(See \cite{Zitomirski,Auerbach,Irkin,DynamicMFA}
and references therein). Although  techniques build upon
boson/fermion representations of spin operators provide an
important insight into the behavior of magnetic systems,
they suffer  from a universal disadvantage:
The su(2) algebra of spin operators and 
the dynamics of the spin system are fully satisfied  only
with exact boson/fermion Hamiltonian and corresponding Hilbert space.
Accordingly, any approximation in the boson/fermion Hamiltonian,
such as  approximate  expression for localized spins or 
some mean-field approximation, destroys the spin nature of 
$(S^{\pm},S^z)$ operators, often in an uncontrolled manner.

%-------------------------------------------------------------------------------------------------------------------
%-------------------------------------------------------------------------------------------------------------------

A different way to deal with magnon-magnon interactions
is based on the equations of motion (EOM) for spin operators \cite{Englert}.
Aside from the fact that the EOM for spin
operators are practically unsolvable for general Heisenberg-type
Hamiltonian, the analysis through EOM gets more involved giving 
the fact that spin operators present basic dynamical variables and 
charge densities at the same time \cite{GHK}.
Despite all that, the method of equations of motion, 
especially in temperature-Green's function (TGF) disguise (See \cite{Kunc}
for a recent review and original references), gained quite a
popularity. There are several reasons for this. First, its predictions
for  thermodynamic properties of Heisenberg-type magnets
 agree with Monte Carlo simulations
\cite{Kunc} and  with experimental results (See e.g. 
\cite{PRB,PRB2,EPJB,SSCCV,SSCTNRPA} and references therein).
Equally important is the flexibility of 
the method making it easily adjustable to systems with
complex \cite{PRBNemci,PRBGFIhle1,PRBGFIhle2,PRBGFIhle3,Semi_2015} or low-dimensional  lattices 
\cite{PRBGFLowD1,PRBGFLowD2,PRBGFLowD3,PRBGFLowD4,PRBGFLowD5,PRBGYabl,PRB1dDM,PRLKYA},
 second and third-neighbor interaction or anisotropies
\cite{J1J2SSC,Plakida2014EPJB,EPJB,PRBGFLongRange,Devlin,PRB,EPJBGFAnis},
which is of great importance when studying real compounds. Also, the TGF method  is 
 recognized as useful in theories of diluted 
magnetic systems \cite{RMPMagSemi, PRBMagSemi}, nuclear spin order in quantum  wires \cite{PRBnucl},
multiferroics models  \cite{PRBMagnEl1,PRBMagnEl2} and even in theories of itinerant electron
systems where Heisenberg Hamiltonian appears as an intermediate effective model
\cite{PRB_itin1,Phil_Mag,Semi_2015}.
The approximations made directly in the equations of motion,
known as the decoupling schemes (DS) in the TGF jargon \cite{Kunc}, 
enable one to solve linearized system of EOM-s thereby
determining the magnon spectrum, correlation functions etc.
By suitably choosing  the parameters  of linearization \cite{KYSOGF,JPSJSOGF}, more or
less satisfactory results may be obtained.
Even though TGF method is extensively exploited  in  contemporary research
(See the  papers quoted earlier in this paragraph),
a thorough discussion on EOM and magnon-magnon interactions seems to be lacking.
This gap in the literature comes from the fact that the nonlinear 
Hamiltonian  describing magnon-magnon interactions need not be introduced
in EOM approach, which makes the comparison with usual boson/fermion representation
theories rather difficult.

The brief sketch  of boson/fermion 
representation and EOM for spin operators
from two preceding paragraphs
 deliberately  points on methodical differences in 
handling   magnon-magnon interactions, as they appear to be
among  frequently used analytical tools and thus cover
a significant part of approximate
treatments of spin systems.
Equally important, however, is their common feature: 
Being based on spin operators and not on the true degrees
of freedom (i.e. magnons), these approximations  tend to obscure the physics 
behind them. 
This is especially true for linearized EOM,
since calculations are conducted without  direct reference
to magnon-magnon interaction operators.

An alternative view on magnons in a  system of localized spins, based upon
symmetry considerations,
is put forward in \cite{PRD}. 
As an extension of the phenomenological Lagrangian method 
\cite{WeinbergSU2,CCWZ1,CCWZ2,Weinberg,
Gasser,Gerber,AnnPhys} to the nonrelativistic systems, the  outcome of
\cite{PRD} is the effective Lagrangian
for the 3D O(3) Heisenberg ferromagnet constructed from the magnon fields
$\pi^1(x)$ and $\pi^2(x)$  transforming nonlinearly under O(3) rotation
group (The subject of nonlinear realizations and  effective Lagrangians  
is reviewed in \cite{WeinbergQTF2}, while texts with special emphasis
on nonrelativistic systems include
 \cite{Burgess,Brauner,RomanSoto,JapanciPRX,JHEP}).
The main advantage in viewing magnons as Goldstone
bosons, accompanying $\text{O}(3) \rightarrow \text{O}(2)$ symmetry breakdown,
is that their dynamics is completely separated from the su(2) algebra
of spin operators entering the Heisenberg Hamiltonian.
Magnons are considered as local coordinates on O(3)/O(2) = $S^2$ and
all information about magnon-magnon interactions 
are encoded in the nonlinear Lagrangian (See Fig. \ref{fig1}). The 
perturbative calculations may be conducted in a systematic manner, 
free of the upper mentioned ambiguities characterizing
spin-operator oriented approaches. The general program of \cite{PRD} is successfully
applied on the calculation of the  free energy
up to two \cite{Hofmann1,Hofmann2} and three loops
\cite{Hofmann3} and is subsequently extended
to low dimensional ferromagnets \cite{Hofmann4,Hofmann5,Hofmann6,Hofmann7}.
If one is  concerned only with low-temperature series of 
the free energy and related quantities, which  are   strongly controlled by 
internal symmetry of the Heisenberg model, the continuum 
field-theoretic methods of \cite{PRD,RomanSoto,Hofmann1,
Hofmann2,Hofmann3,Hofmann4,Hofmann5,Hofmann6,Hofmann7} will suffice. However,
the comparison with experimental data and other spin-operator-based methods
often requires for the discrete symmetry of the lattice to be included also.
This is the case, for example, with magnon energies whose dispersion deviates 
from simple $\bm k^2$ law due to the lattice anisotropies.
The symmetry of the lattice may be incorporated in the effective
field description by the method of \cite{PaperAnnPhys} (See also \cite{RomanSoto}),
resulting in an effective theory which describes ferromagnet with the
help of lattice magnon fields. This means that the internal O(3)
symmetry is implemented through the usual nonlinear  realization (equivalently,
symmetry may be realized linearly on the magnon fields, but with an
additional nonlinear constraint), while the lattice symmetries enter
via  adequately chosen regulator. At the same time, 
the use of spatial lattice  singles out  canonical formalism
in constructing
quantum field-theoretic description of a ferromagnet.
Beside the fact that the Hamiltonian lattice
theory and the functional Lagrangian method 
with dimensional regularization
yield the same results, as it was shown in \cite{PaperAnnPhys},
few other important issues favor canonical over functional integral formalism
in this particular case.
First, with the lattice magnon Hamiltonian at hand, 
 a comparison with standard methods based on
boson representations of the spin operators \cite{Zitomirski,Auerbach,Irkin,KaganovChubukov,
Borovik} becomes a much easier task. Second, EOM method 
(i.e. two-time temperature Green's function theory) which is be systematically
reexamined in present paper, is intimately related to the spin operator
Hamiltonian through the definitions of thermodynamic averages.
Also, the procedure of quantization itself is insightful regarding
how magnon-magnon interactions  may appear in an  effective field theory. 
And finally, interacting Hamiltonian for ferromagnetic magnons, rather than the general
form of interacting Lagrangian,  reveals certain simplifications in
loop calculations \cite{PaperAnnPhys} if a Lagrangian contains
the Wess-Zumino term.

The lattice Hamiltonian theory of magnon fields provides us with 
an efficient tool for studying effects of  magnon-magnon interactions
in O(3) ferromagnets, and likewise, the  interactions induced by
linearizations in  EOM approach.  This is an extremely important point since,
according to the current understanding \cite{Kunc},
the quality of DS can be judged only \emph{a posteriori}.
Also, it is difficult to distinguish between the true effects
of magnon-magnon interactions induced by linearizations of EOM
from pure artifacts of TGF formalism. All this can lead to
incorrect interpretation of results obtained using EOM for spin operators.
Previous remarks bring us to the main purpose
of the present article.
By explicitly identifying types of magnon-magnon interactions that
actually constitute some of the well known DS and connecting
them with internal symmetry of the O(3) ferromagnet, we  put EOM approach
in broader context of interacting spin waves and effective Lagrangians
thereby connecting it with recent developments in these fields.
At the same time  we clarify some misinterpretations
of EOM method  in the case of O(3) ferromagnets. 
Quite generally, the effects of magnon-magnon interactions
can be inferred %(read off) 
either from   low-temperature expansions of the
free energy and related thermodynamic quantities or from
renormalized magnon energies, as corresponding results
in linear spin wave  (LSW) approximation are well known \cite{PaperAnnPhys}.
Since the heart of EOM method lies in determination of the
renormalized magnon energies, we shall concentrate on 
finding appropriate effective Lagrangian that yields, through the
perturbation theory, desired magnon self-energy.
This will allow us to determine the true nature of some well known
linearizations from the physical point of view, i.e. relying only
on   magnon-magnon interactions and internal symmetry of
the Heisenberg model.

The perturbation theory for lattice magnon fields shares
many features with the standard effective field theory. 
The most important distinction concerns calculation details,
as the spatial lattice replaces continuum field theoretic setting.
Section \ref{LattLaplSect} summarizes main definitions and relations
needed for loop calculations which are presented in Sec. \ref{SecLMH}
and \ref{SecSOA}. Specifically, it is shown in Sec. \ref{SecLMH} 
that the random phase approximation (RPA) and the Callen approximation (CA)
can  be understood as  interaction theories  for type B magnons
(see \cite{JapanciPRL,JapanciPRX} for A/B classification of 
Goldstone fields). Finally, Sec. \ref{SecSOA} describes
the second-order approximation of Kondo and Yamaji for one dimensional
ferromagnet as a perturbation theory for type A magnons.
%on  a chain lattice. 
Further justification of the lattice magnon field formalism
comes from comparison with Quantum Monte Carlo simulation
[sections \ref{QMCB} and \ref{QMCA}] which, at the same time, supports
some of our conclusion regarding RPA, CA and KYA.

\section{Lattice Laplacians and magnon-magnon interactions} \label{LattLaplSect}

Conventional effective field theory (EFT)
combines derivative and loop expansions to set up perturbative calculations
to given order in some energy scale \cite{WeinbergQTF2,Burgess,Brauner}.
The relevant scale for ferromagnets is determined by temperature
and EFT enables one to systematically calculate free energy
in powers of $T$ \cite{Gerber,Hofmann1,Hofmann2,Hofmann3,Hofmann4,Hofmann5,Hofmann6,Hofmann7}.
Yet, as emphasized in the Introduction, some issues concerning
ordered magnets are adequately treated by the perturbation 
theory for lattice magnon fields which preserves the
full symmetry of  Heisenberg Hamiltonian. 
This approach, adjusted for O(3) 
ferromagnets and systematically exposed in following sections, 
includes couplings of neighboring sites through the lattice Laplacian.
For general lattice  it is defined by \cite{LatLapl}
%------------------------------------------------------------------------------------------------------------------------
\bea
\hspace*{-0.4cm}\nabla^2 \phi(\bm x)   = \frac{2D}{Z_1 |\bm \lambda|^2} 
 \sum_{  \{ \bm \lambda \}  } \Big{[}\phi(\bm x + \bm \lambda) 
-  \phi(\bm x)\Big{]},
\label{OpIzLapA}
\eea
%------------------------------------------------------------------------------------------------------------------------
where $\bm x$ denotes a lattice site, $D$ is the dimensionality
of  spatial lattice and  $\{ \bm \lambda \}$ are vectors
connecting each lattice site with its $Z_1$ nearest neighbors. The
definition (\ref{OpIzLapA}) can be extended in an obvious manner to
next-nearest neighbor  couplings.
The plane wave solutions, corresponding to free magnons,
are built with eigenvalues of $\nabla^2$
%-------------------------------------------------------------------------------------------------------------------
\bea
&&\nabla^2 \exp [\i \bm k \cdot \bm x]  =  
-\frac{2 D}{|\bm \lambda|^2} [1-\gamma_{D}(\bm k)] \exp [\i \bm k \cdot \bm x] 
\label{kSq} \\
& \equiv &  - \widehat{\bm k}^2 \exp [\i \bm k \cdot \bm x], \;\;\;\;\;\;
 \gamma_{D}(\bm k) = Z_1^{-1} \sum_{\{ \bm \lambda \}} \exp [\i \bm k \cdot \bm \lambda].
 \nonumber
\eea
%-------------------------------------------------------------------------------------------------------------------
Depending the presymplectic structure
\cite{JapanciPRX,JapanciPRL}, Goldsone bosons may be classified as
type A or  B. In the present context, magnons of
type A  posses free energies proportional to
 $\widehat{\bm k}$  and those of type B  to $\widehat{\bm k}^2$.
As a consequence of derivative couplings, the loop integrals
generally contain eigenvalues combining
loop and external momenta, 
$\widehat{\bm p - \bm q}\;^{2}$. These can be simplified according to
%------------------------------------------------------------------------------------------------------------------------
\bea
\hspace*{-0.45cm}\int_{\bm q} \Lan n_{\bm q} \Ran_0 \; \widehat{\bm p - \bm q}\; ^{2} = 
\int_{\bm q} \Lan n_{\bm q} \Ran_0 \left[\widehat{\bm p}^{\;2} + \widehat{\bm q}^{\;2}  
- \frac{|\bm \lambda|^2}{2 D} \widehat{\bm p}^{\;2} \; \widehat{\bm q}^{\;2} \right]
\label{SimplInt}
\eea
%------------------------------------------------------------------------------------------------------------------------
where $ \Lan n_{\bm p}\Ran_0$ denotes Bose distribution for free magnons
and
%------------------------------------------------------------------------------------------------------------------------
\bea
\int_{\bm q} \equiv  \int_{\text{IBZ}} \frac{\d^D \bm q}{(2 \pi)^D}. \label{Int1BZ}
\eea
%------------------------------------------------------------------------------------------------------------------------
The focus in Sec. \ref{SecSOA} is on one dimensional ferromagnets 
described by an effective theory that includes couplings of lattice magnon fields
spanning beyond nearest  neighbors. To
make presentation in Sec. \ref{SecSOA} more transparent, we shall denote with %$\nabla^2$
$\nabla^2_{(2)}$ the Laplacian for  second neighbor
couplings on a chain. It is defined by
%-------------------------------------------------------------------------------------------------------------------
\bea
\nabla^2_{(2)} \e^{\i q x} & = &- \frac{1}{2 a^2}\left[ 1-\gamma(2 q) \right]\e^{\i q x}
\equiv - \widehat{\;2 q\;}^2 \e^{\i q x}, \label{LattLapl_1&2} 
\eea
%-------------------------------------------------------------------------------------------------------------------
so that
%-------------------------------------------------------------------------------------------------------------------
\bea
\widehat{\;2 q\;}^2 = {\widehat{q}^2} \left[ 1 - \frac{a^2}{4} \widehat{q}^2 \right]
\label{LattLapl_1&2_b} 
\eea
%-------------------------------------------------------------------------------------------------------------------
and $\gamma(q)$ is defined in (\ref{kSq}) for $D=1$. An analogue of (\ref{SimplInt})
for $\nabla^2_{(2)}$ reads
%------------------------------------------------------------------------------------------------------------------------
\bea
\hspace*{-0.45cm}\int_{q} \Lan n_{q} \Ran_0 \hspace{-0.2cm} 
\reallywidehat{2   p - 2  q}\hspace{-0.2cm}^{2} & = & 
\int_{q} \Lan n_{q} \Ran_0  \left[\widehat{\; 2 p \;}^{\;2}  +  \widehat{\; 2 q \;}^{\;2}   \right.  \nonumber \\
&-& \left. 2 a^2 \widehat{\; 2 p \;}^{\;2} \; \widehat{\; 2q \;}^{\;2} \right].
\label{SimplInt2}
\eea
%------------------------------------------------------------------------------------------------------------------------
Equations from this section will be used throughout the paper for
loop calculations, frequently  without  direct reference.

%===============================================================================================================================================
\section{Perturbation theory for Type B magnons}\label{SecLMH}
%===============================================================================================================================================

We introduce lattice magnon Hamiltonian and accompanying
diagrammatic rules for type B magnons \cite{JapanciPRL,JapanciPRX}
in this section. They will serve as
a basic for the systematic investigation of several approximate
treatments of magnon-magnon interactions in O(3) ferromagnes
for $D \geq 3$.

%===============================================================================================================================================
\subsection{Lattice magnon Hamiltonian}\label{SubecLMH}
%===============================================================================================================================================

The starting point in an effective field description of a ferromagnet
is the classical Lagrangian \cite{PRD} (See also \cite{Jevicki,Klauder,Wiese})
of the unit vector field $U^i:=[U^1,U^2,U^3]^{\text T}
\equiv [\bm \pi (x),U^3(x)]^{\text T}$ transforming linearly under O(3)
%-------------------------------------------------------------------------------------------------------------------
\bea
\L_{\text{eff}} & = & \Sigma \frac{\partial_t U^1 U^2 - \partial_t U^2 U^1}{1+U^3}
-\frac{F^2}{2} \partial_{\alpha} U^i \partial_{\alpha} U^i  \nonumber \\
& + & \Sigma \mu H U^3. \label{EffLagr}
\eea
%-------------------------------------------------------------------------------------------------------------------
$\Sigma = N S/V$ is the spontaneous magnetization per unit volume at $T = 0$K,
$F$ is a constant, to be fixed latter, $H$ is the external field and $\alpha$
refers to the spatial coordinates. We also  recall that the Lagrangian (\ref{EffLagr})
collects all magnon-magnon interactions up to $\bm p^2$. 
Due to the  nonvanishing charge densities in the ground state and the non-Abelian
character of the O(3) rotational group, the effective Lagrangian
contains   appropriate Wess-Zumino (WZ) term. Being linear in
temporal derivatives, WZ term is responsible for nonrelativistic
dispersion of ferromagnetic magnons \cite{PRD,WenZee,BraunerMoroz}
and (\ref{EffLagr})
describes the system of type-B Goldstone bosons \cite{JapanciPRL}. In contrast to this 
well known fact, its influence in perturbation theory, i.e in loop corrections
to the magnon self-energy and the ferromagnet free energy, is not fully appreciated.
The reason for this is that WZ term is practically invisible in approaches
based on boson representations, dominating the literature on magnon-magnon
interactions \cite{Zitomirski,KaganovChubukov,Borovik}, 
since it is hidden inside the commutation relations of spin operators.

As outlined  in the Introduction, we shall employ canonical
formalism in constructing the quantum theory. We first note that the
field describing  physical magnons is  complex \cite{PRD, Hofmann1,JapanciPRL},
$\psi (x) = \sqrt{\Sigma/2}[ \pi^1(x) + \i \pi^2 (x)]$, since the equation governing
their dynamics is of a Schr\"{o}dinger type.
The standard canonical prescription \cite{WeinbergQTF1} then yields the 
canonical momentum in the form
$\Pi = 2\i \psi /[1+U^3]$, with  $U^3 = \sqrt{1-(2/\Sigma) \psi^\dagger \psi}$. 
This path towards consistent quantum theory is of course 
completely legitimate, but nonlinear connection between
$\psi^\dagger$ and $\Pi$ sends important part of the magnon-magnon
interactions from the Hamiltonian $\H (\psi, \psi^\dagger)$
to the canonical commutation relations, $[\psi(\bm x), \Pi(\bm y)]
= \i \delta(\bm x-\bm y)$. The quantum field theory of this sort is 
strongly reminiscent of the initial,  spin-operator oriented, approach
and is in fact something we would like to avoid.
Fortunately, all complications  concerning canonical commutaion
relations  can be circumvented   by noting that terms with single temporal
derivative may be eliminated from the Lagrangian (\ref{EffLagr})
with the help of the equation of motion for $\bm U(\bm x,t)$ \cite{WeinbergQTF1} (See
also \cite{Hasenfratz_EOM_COV}).
In this manner, we obtain the free magnon Lagrangian (bilinear in
magnon fields $\pi^1$ and $\pi^2$) and nonlinear part describing their interaction
\cite{PaperAnnPhys}. Canonical quantum interaction theory
may be constructed starting from this Lagrangian.
For the present purposes, we shall be needing terms
up to and including eight magnon field operators. An straightforward
calculation yields redesigned Hamiltonian \cite{PaperAnnPhys}. This Hamiltonian,
however, includes only magnon-magnon interactions of the
order $\bm p^2$. It is well known \cite{Weinberg,Gerber,Gasser2,AnnPhys,Hofmann2,Hofmann3}
that consistent loop expansion must include interaction terms of
higher order in magnon momenta. Following examples
from Lorentz-invariant chiral perturbation theory \cite{Smilga,Lewis},
we include higher order terms in momentum, i.e. in spatial
derivatives, by putting the Hamiltonian on the (spatial) lattice.
We find terms with four and eight magnon field operators
%-------------------------------------------------------------------------------------------------------------------
\bea
H_{\text{eff}} & = & H_0 + H_{\text{int}}, \label{HamEffLatt0} \\
H_0 &  = & -\frac{1}{2 m_0} v_0 \sum_{\bm x}
 \psi^\dagger(\bm x) \nabla^2 \psi(\bm x), \nonumber \\
m_0 & = & \frac{\Sigma}{2 F^2} ,       \nonumber \\
H_{\text{int}} & = & H_4^{(a)}+H_4^{(b)} + H_8^{(a)}+H_8^{(b)}, \nonumber
\eea
%-------------------------------------------------------------------------------------------------------------------
where $v_0$ denotes the volume of the unit cell and
%-------------------------------------------------------------------------------------------------------------------
\bea
H_4^{(a)} & = &\frac{F^2}{8}  v_0 \sum_{\bm x} \bm \pi^2(\bm x)  \bm \pi(\bm x) \cdot \nabla^2 \bm \pi(\bm x)
\label{HamEffLatt1}, \\
H_4^{(b)} & = & -\frac{F^2}{8}  v_0 \sum_{\bm x} \bm \pi^2(\bm x) \nabla^2 \bm \pi^2(\bm x), \nonumber %\\
\eea
%----------------------
\bea
H_8^{(a)} & = & -\frac{F^2}{128}  v_0 \sum_{\bm x} \left[\bm \pi^2(\bm x) \right]^3
 \bm \pi(\bm x) \cdot \nabla^2 \bm \pi(\bm x), \nonumber \\
H_8^{(b)} & = &\frac{F^2}{128}  v_0 \sum_{\bm x} \left[\bm \pi^2(\bm x) \right]^2
 \nabla^2 \left[ \bm \pi^2(\bm x) \right]^2  \label{HamEffLatt2},
\eea
%-------------------------------------------------------------------------------------------------------------------
We have set $H=0$ for the time being. 
In what follows,  $\psi$ will always be written to 
the right in expressions like $\bm \pi \cdot \bm \pi$.
We also note that $\Sigma  = S/v_0$ in a lattice theory,
with $S$ denoting the magnitude of localized spins. The LSW spectrum
is fully recovered if we choose $F^2/\Sigma = J S Z_1 |\bm \lambda|^2/(2 D)
= 1/(2 m_0)$ [see equation (\ref{Omega0}) bellow].
This simple connection between parameters of the effective
theory and those of original Heisenbeg model is a consequence
of the ferromagnetic ground state.

Some comments on interacting magnon Hamiltonian (\ref{HamEffLatt0})-(\ref{HamEffLatt2}) are in order.
We observe that magnon-magnon interactions
come in two  categories, distinguished by superscripts
$a$ and $b$ in (\ref{HamEffLatt1})-(\ref{HamEffLatt2}). While 
$H_4^{(b)}$ and $H_8^{(b)}$  are typical interaction terms
for theories of unit vector fields, those containing 
operator $\bm \pi(\bm x) \cdot \nabla^2 \bm \pi(\bm x)$, 
namely $H_4^{(a)}$ and $H_8^{(a)}$, describe the magnon-magnon
interactions originating in WZ term. Thus, they   are specific
to the ferromagnetic system and careful treatment
of the magnon-magnon interactions arising in  WZ term is mandatory,
if the correct description of a ferromagnet is to be expected.
We shall come back to this point several times in the text.
It can also be  seen from (\ref{HamEffLatt0}) and (\ref{HamEffLatt1})
that $H_0 + H_4^{(a)} + H_4^{(b)}$ is quite similar to 
the Dyson-Maleev Hamiltonian \cite{Dyson1,Dyson2,Maleev}.
Both the Dyson's construction and the Maleev 
boson representation \footnote{Originally, Dyson obtained Hamiltonian
 from the requirement
that certain combination of boson operators has the same effect
on ideal spin-wave states as the  Heisenberg Hamiltonian does
 on spin-wave states (See \cite{Dyson1} for Dyson's  definitions
of the spin-wave states). The Dyson Hamiltonian is also a 
direct consequence of the Dyson-Maleev \cite{Maleev} boson representation
of spin operators.} yield magnon Hamiltoninan equivalent
to $H_0 + H_4^{(a)} + H_4^{(b)}$, but only
the method based on the effective Lagrangian (\ref{HamEffLatt0})
traces    $\bm \pi^2(x)\bm \pi(\bm x) \cdot \nabla^2 \bm \pi(\bm x)$
interaction back to
the WZ term. Finally, the   Hmiltonaian given in (\ref{HamEffLatt0})
and (\ref{HamEffLatt1})-(\ref{HamEffLatt2}), being the Hamiltonian of true
Goldstone fields,  makes the weakness of magnon-magnon 
interactions at low momenta rather obvious. 
Thus, $1/S$ expansion, a usual
starting point in theories based on boson representations
and \emph{a priori} justification for the weakness of magnon-magnon interactions
\cite{Zitomirski,Auerbach,KaganovChubukov,Borovik} is, in fact, unnecessary.
The magnons interact weakly at low momenta  irrespective of the spin magnitude $S$,
and this is what actually makes the perturbation theory  well defined \footnote{To
be precise, the weakness of magnon-magnon interactions is a
characteristic of  spin systems with collinear order, see
\cite{Zitomirski}} 
even for $S=1/2$.

%===============================================================================================================================================
\subsection{Diagramatic description of magnon-magnon interactions}\label{DiagDesc}
%===============================================================================================================================================

Before proceeding  with concrete  perturbation theory calculations, we
pause to introduce a useful variant of Feynman diagrams \cite{PaperAnnPhys}. 
It is particularly suited to the magnon Hamiltonian
(\ref{HamEffLatt0}), but it can be easily adapted to other scalar 
field theories with derivative couplings.
This method will allow us to systematically keep  track of individual
contributions in various magnon-magnon processes, which will
turned out to be of great importance when comparing
magnon perturbation theory with approximations 
in the equations of motion.
To account for the eight-magnon vertices, we slightly modify the notations
used in \cite{PaperAnnPhys}.

We shall denote with
$\begin{array}{r}
 {\includegraphics[scale=1.1]{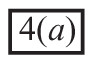}}
\end{array}$, 
$\begin{array}{r}
 {\includegraphics[scale=1.0]{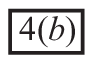}}
\end{array}$,
$\begin{array}{r}
 {\includegraphics[scale=1.0]{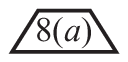}}
\end{array}$ and $\begin{array}{r}
 {\includegraphics[scale=1.0]{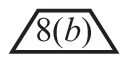}}
\end{array}$ the vertices corresponding to interaction
terms from (\ref{HamEffLatt1})-(\ref{HamEffLatt2}).
Depending on the structure
of the vertex, lattice Laplacians may act on one, two
or four magnon propagators. Apart from simplest,  one-vertex 
diagrams,  higher order corrections will always contain
at least two lattice Laplacians acting on propagators.
The calculations soon become quite involved, since the number
of propagators, as well as the number of lattice Laplacians
acting upon them, increases.
To make them feasible, we associate with each vertex
a color so that the vertex and the
propagators affected by its lattice Laplacian are  drawn
using that color.
The rest of the lines are simply  drawn  black.
We label each propagator with $D+1$ dimensional momentum
$k := [\bm k, \omega_n]^{\text T}$ and the colored ones 
also carry  $-\widehat{\bm k}^2$. 
Here $-\widehat{\bm k}^2$ denotes the eigenvalue of the lattice
Laplacian, defined in (\ref{kSq}),
and  $\omega_n = 2 \pi n T$ are the Matsubara frequencies.
When colored line passes through the vertex, the momentum in $-\widehat{\bm k}^2$
corresponds to algebraic sum of the incoming and outgoing
momenta. If more than one Laplacian acts on a certain propagator,
the line will carry one color for each vertex. We shall adopt the following convention
on colored propagators belonging to a closed loop: If the lattice Laplacian 
of $\begin{array}{r}
 {\includegraphics[scale=1.1]{Diag_1.eps}}
\end{array}$ or 
$\begin{array}{r}
 {\includegraphics[scale=1.0]{Diag_3.eps}}
\end{array}$   acts on a magnon propagator closing a
loop around it, the line will be drawn half-colored.
When the Laplacian comes from $\begin{array}{r}
 {\includegraphics[scale=1.1]{Diag_2.eps}}
\end{array}$ or 
$\begin{array}{r}
 {\includegraphics[scale=1.0]{Diag_4.eps}}
\end{array}$, the line will be drawn in full color if 
the Laplacian acts on both contracted magnon field operators.
Otherwise, the line will also be drawn half-colored.
With this conventions we generalize the method of \cite{PaperAnnPhys}
to interacting terms with more than four magnon operators.
The diagrams with colored lines and vertices will be 
referred to  as the colored contractions.
Later, in Sec. \ref{SecSOA}, we shall allow more than one
Laplacian per vertex.
Lastly, the magnon propagator and the free one-magnon energies
for type B magnons are given by
%-------------------------------------------------------------------------------------------------------------------
\bea
\hspace*{-0.2cm}D(\bm x - \bm y, \tau_x - \tau_y) & \equiv & 
\Lan \mbox{T} \left\{ \psi(\bm x,\tau_x) \psi^\dagger (\bm y,\tau_y) 
 \right\} \Ran_0  \label{Prop} \\
& = &  \frac 1\beta \sum_{n = -\infty}^\infty 
\int_{\bm q} \frac{\e^{\i \bm q \cdot (\bm x - \bm y) - \i
 \omega_n (\tau_x-\tau_y)}}{\omega_0(\bm q) - \i \omega_n}, \nonumber 
\eea
%-------------------------------------------------------------------------------------------------------------------
%-------------------------------------------------------------------------------------------------------------------
\bea
\omega_0(\bm q)  =  \frac{\widehat{\bm q}^2}{2 m_0},  && \hspace{0.5cm}
 \label{Omega0}
\eea
%-------------------------------------------------------------------------------------------------------------------
and  $\beta^{-1} \sum_n [\omega_0({\bm p})-\i \omega_n]^{-1} = \Lan n_{\bm p}\Ran_0$,
where $ \Lan n_{\bm p}\Ran_0$ denotes Bose distribution for free magnons.

As an explicit example, consider a three-loop correction to the 
magnon self-energy arising from $H_8^{(b)}$. Formally, it is represented by 
%------------------------------------------------------------------------------------------------------------------------
\bea
\Sigma_8^{(b)}(\bm k) =
\begin{array}{l}
\vspace{0.1cm}\includegraphics[scale=0.75]{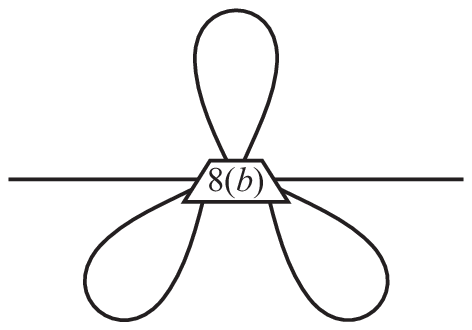}
\end{array}.
\eea
%------------------------------------------------------------------------------------------------------------------------
Several colored contractions contribute to this
particular diagram. These are 
\begin{widetext}
%------------------------------------------------------------------------------------------------------------------------
\bea
\begin{array}{l}
\includegraphics[scale=0.8]{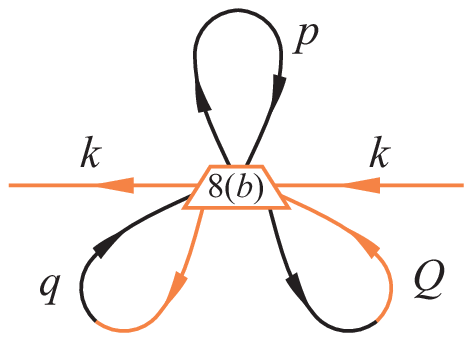}
\end{array}&  + &
\begin{array}{l}
\includegraphics[scale=0.8]{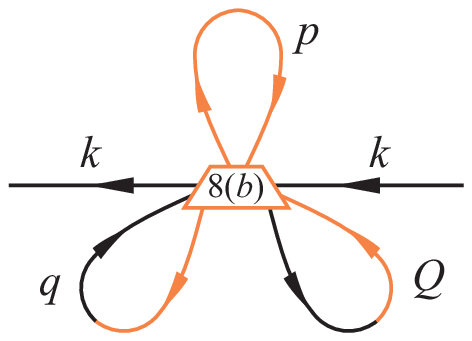}
\end{array}
=  -\frac{4}{S^3} \frac{\Lan n_{\bm x}\Ran}{2 m_0} v_0^2 \int_{\bm Q, \bm q}
\Lan n_{\bm Q}\Ran_0 \Lan n_{\bm q}\Ran_0 \; \widehat{\bm Q - \bm q}\;^2,  \label{Sigma8b-1}%\\
\eea
\bea
\begin{array}{l}
\includegraphics[scale=0.8]{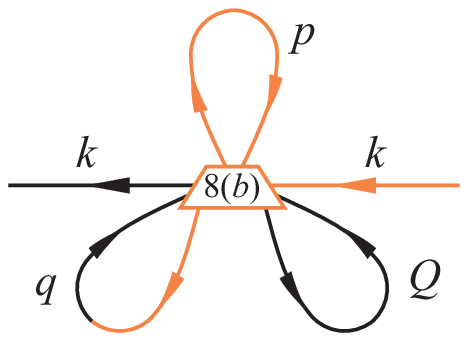}
\end{array}
 & + &
\begin{array}{l}
\includegraphics[scale=0.8]{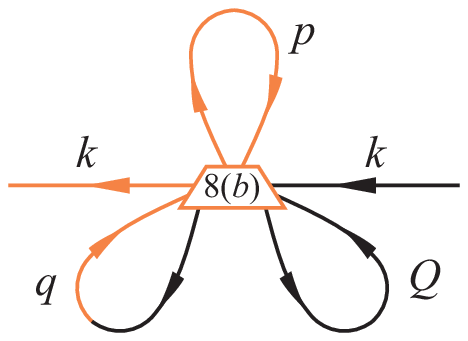}
\end{array}
 =  -\frac{4}{ S^3} \frac{\Lan n_{\bm x}\Ran^2}{2 m_0} v_0 \int_{\bm q}
 \Lan n_{\bm q}\Ran_0 \; \widehat{\bm k - \bm q}\;^2 , \label{Sigma8b-2} %\\
 \eea
 \bea
\begin{array}{l}
\includegraphics[scale=0.8]{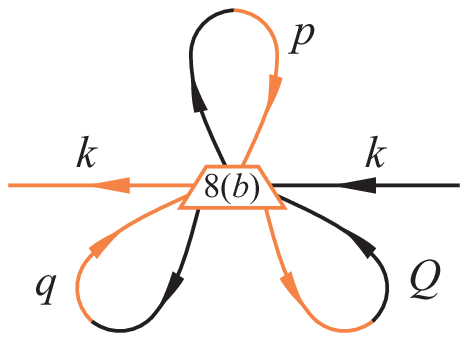}
\end{array}
 & + &
\begin{array}{l}
\includegraphics[scale=0.8]{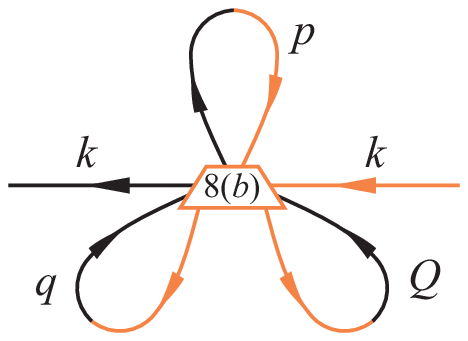}
\end{array}
 =  -\frac{2}{ S^3} \frac{v_0^3}{2 m_0}  \int_{\bm q, \bm p, \bm Q}
 \Lan n_{\bm q}\Ran_0  \Lan n_{\bm p}\Ran_0
  \Lan n_{\bm Q}\Ran_0 \;
{\overset{{\begin{array}{l}
\includegraphics[scale=0.8]{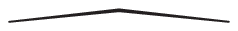}
\end{array}}}
{\bm p + \bm q - \bm Q - \bm k}} {\hspace{0.01cm}^{2}} 
\nonumber \\
\label{Sigma8b-3}
\eea 
%------------------------------------------------------------------------------------------------------------------------
\end{widetext}
since two colored loops  closed around single vertex,
as well as single colored closed loop with two colored external
lines, vanish due to momentum conservation:
%------------------------------------------------------------------------------------------------------------------------
\bea
\begin{array}{l}
\vspace{-0.3cm}\includegraphics[scale=0.9]{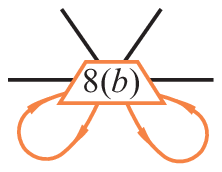}
\end{array} =  \begin{array}{l}
\vspace{0.22cm}\includegraphics[scale=0.9]{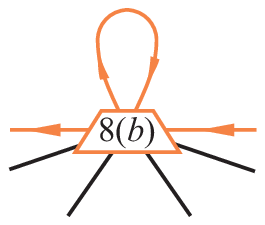}
\end{array}
=0 .     %\begin{array}{l}
\eea
%------------------------------------------------------------------------------------------------------------------------
This is the analog of vanishing one-loop colored
diagram of $H_4^{(b)}$ \cite{PaperAnnPhys}. In the 
equations above  
$\Lan n_{\bm x}\Ran = v_0\int_{\bm q} \Lan n_{\bm q}\Ran_0$
is the mean number of magnons per lattice site.
Further examples, including three-loop corrections
to self-energy  from $H_4^{(a)}$ and $H_4^{(b)}$, can be
found in \cite{PaperAnnPhys}. Also, following \cite{PaperAnnPhys},
we omit combinatorial factors in front of diagrams 
and include them directly  in corresponding equations.
For example, each of colored contractions from equation (\ref{Sigma8b-1})
is accompanied by a combinatorial factor of 16, which
is just the number of permutations of black and colored propagators
which leave the diagram unchanged.
We note one, more or less obvious, fact
concerning one-vertex diagrams:
the number of closed loops equals the number of  $\Lan n_{\bm q} \Ran_0$'s.
This can easily be checked by inspection of the equations (\ref{Sigma8b-1})-(\ref{Sigma8b-3}).

%===============================================================================================================================================
\subsection{Two-loop correction to the magnon self-energy}\label{2LoopSE}
%===============================================================================================================================================

To calculate the two-loop correction to the magnon self-energy, one must
include magnon-magnon interactions described by $H_4^{(a)}$ and $H_4^{(b)}$.
A direct computation  shows that one-vertex diagrams, as well as
two-vertex diagrams with both external lines attached to the same vertex,
simply renormalize the magnon mass \footnote{Ferromagnetic magnons are
massless from the point of the view of the Goldstone theorem, but 
giving the nonrelativistic form of the dispersion (See (\ref{Omega0})), it
is convenient to refer to $m_0$ and $m(T)$ as  magnon mass} ($m_0 \rightarrow m(T)$),
while the "sunset" diagrams (each vertex carrying single external
and three internal lines)
change the geometry of the magnon dispersion. The result  is 
\cite{PaperAnnPhys}
\begin{widetext}
%------------------------------------------------------------------------------------------------------------------------
\bea
\Sigma(\bm k, \i \omega_n) & = & \frac{\widehat{\bm k}^2}{2 m_0} \left[A(T) + A(T) B(T)\right] 
+
\frac{\widehat{\bm k}^2}{2 m_0}  \frac{1}{2 S^2} 
\left(  \frac{|\bm \lambda|^2}{2 D}  \right)^2 \frac{v_0^2}{2 m_0 } \int_{\bm p,\bm q}
F^{\bm k}_{\bm p, \bm q}\left(\i \omega_n \right) 
 \widehat{\bm q}^{\;2}\left( \widehat{\bm p}^{\;2} + \widehat{\bm k}^{\;2} \right)
\left( \widehat{\bm q}^{\;2}   - \widehat{\bm p - \bm q}^{\;2} \right), \nonumber \\
F^{\bm k}_{\bm p, \bm q}\left(\i \omega_n \right) & = & 
\frac{
\Lan n_{\bm p} \Ran_0 [1+\Lan n_{\bm q} \Ran_0 + \Lan n_{\bm k + \bm p - \bm q} \Ran_0]
-\Lan n_{\bm q} \Ran_0 \Lan n_{\bm k + \bm p - \bm q} \Ran_0}
{\omega_{0}(\bm k + \bm p - \bm q)- \omega_{0}(\bm p) + \omega_{0}(\bm q) - \i \omega_n}, \label{2LoopSEM} \\
A(T) & = & \frac 1 S \frac{|\bm \lambda|^2}{2D} \; v_0 \int_{\bm q} \Lan n_{\bm q} \Ran_0 \widehat{\bm q}^{\;2},
\nonumber \\
B(T) & = & \frac{1}{S} 
 \frac{1}{2 m_0 T} \;
\frac{|\bm \lambda|^2}{2 D}\;  v_0 \int_{\bm p} \frac{ \Lan n_{\bm p} \Ran_0 [\Lan n_{\bm p} \Ran_0 +1]}{T} 
\left[ \widehat{\bm p}^{\;2}  \right]^2, \nonumber
\eea
%------------------------------------------------------------------------------------------------------------------------
\end{widetext}
where $F^{\bm k}_{\bm p, \bm q}(\i \omega_n)$ comes from the Matsubara summation
in the sunset diagram, and $\Lan n_{\bm p} \Ran_0[\Lan n_{\bm p} \Ran_0 +1]/T$
is induced by two-loop diagrams in which both external lines are
attached to the same vertex (See \cite{PaperAnnPhys}).
Renormalized magnon energies are now easily found as 
$\omega_{\text{2loop}}(\bm k)  =  \omega_0(\bm k) - \delta\omega_{\text{2loop}}(\bm k)$,
with $\delta\omega_{\text{2loop}}(\bm k) =   \lim_{\delta \rightarrow 0}      
\text{Re} \Sigma(\bm k, \omega_{0}(\bm k)+ \i \delta)$. It is seen that
magnons,  as  pions in Lorentz-invariant theories \cite{Smilga,StrangeMass},
remain gapless at two loop. In the rest of this section we shall
exploit perturbation theory to test certain approximate
treatments of magnon-magnon interactions against two-loop result (\ref{2LoopSEM}).
The calculation that led to (\ref{2LoopSEM}) will be referred to as true magnon
interaction theory, since it preserves spin-rotation symmetry
up to and including two-loop corrections to the magnon self-energy.

%===============================================================================================================================================
\subsection{The Random Phase Approximation}\label{SecRPA}
%===============================================================================================================================================

This well known approximation \cite{Kunc} is usually described 
as the one in which correlations between $S^z$ and
$S^{\pm}$ operators from adjacent sites are neglected. 
Though it is most frequently discussed within TGF formalism,
it can be easily obtained by replacing $S_{\bm x}^z(t)$ with site-independent
average $\Lan S^z \Ran$ in the equations of motion
and commutation relations  for $S_{\bm x}^\pm(t)$.
The net effect of this approximation is the renormalization of  magnon mass,
$m_0 \rightarrow m_0 \times S/\Lan S^z \Ran$, leading to 
incorrect description of thermodynamics at low temperatures
(the presence of  so-called spurious $T^3$ term in low-temperature
series for spontaneous magnetization).
It should be obvious from renormalized magnon spectrum and low-temperature
series of spontaneous magnetization  that RPA transforms 
localized interacting spins into a system of lattice bosons, as
one can introduce Schr\"{o}dinger field operators 
$[\psi(\bm x,t), \psi^\dagger(\bm y,t)]  =  v_0 \Delta(\bm x - \bm y)$,
in terms of which
%===============================================================================================================================================
\bea
S^+_{\bm x}(t) \overset{\text{RPA}}{=}  \sqrt{2 \Lan S^z \Ran} \psi (x),  \;\;\;\;&& 
S^-_{\bm x}(t)  \overset{\text{RPA}}{=}  \sqrt{2 \Lan S^z \Ran} \psi^\dagger (x), \nonumber \\
\omega_{\text{RPA}}(\bm k)  =  \frac{\widehat{\bm k}^2}{2 m_{\text{RPA}}},\;\;\;\; &&
 m_{\text{RPA}} = m_0 \frac{S}{\Lan S^z \Ran},\label{PsiRPA}
\eea
%===============================================================================================================================================
and the  RPA description of O(3) HFM formally corresponds to the free field solution of this
bose theory. For example, the correlation function $\Lan S^-_{\bm x} S^+_{\bm x} \Ran$
can be written as $\Lan S^-_{\bm x} S^+_{\bm x} \Ran_{\text{RPA}}=2  \Lan S^z  \Ran  v_0
\Lan \psi^\dagger(\bm x) \psi(\bm x) \Ran_{\text{RPA}}=2  \Lan S^z  \Ran v_0
\int_{\bm k} \Lan n_{\bm k} \Ran_{\text{RPA}}$, where  
$\Lan n_{\bm k} \Ran_{\text{RPA}}$ denotes the Bose distribution for 
magnons with energies $\omega_{\text{RPA}}(\bm k)$. This 
agrees with standard TGF theory \cite{Kunc}.
Much less trivial is the question of magnon-magnon interactions
induced by RPA.

After original papers of Tyablikov and Bogolyubov, a number of authors 
\cite{Stinchcombe,VLP1,VLP2,Fishman,KineziRPA} 
succeeded  in obtaining the same results using different
techniques.  Where they all had  failed  is to connect RPA results
with magnon-magnon interactions and internal symmetry of the model,
since all those works  (including the original one of 
Tyablikov and Bogolyubov) are spin-operator oriented. In particular,  
the perturbation theory in \cite{VLP1,VLP2,KineziRPA}
is built on convenient MFA and fermion representation of spin
operators, respectively, while \cite{Stinchcombe,Fishman} develop
linked cluster and $1/Z$ expansion.

It is only recently \cite{PaperAnnPhys} suggested  that 
RPA can be obtained in physically more transparent manner,
using the perturbation theory for
lattice magnon fields.  The  approximation in question
is the replacement of $H_{\text{eff}} = H_0+H_4^{(a)} + H_4^{(b)}$ with
$H_{\text{RPA}} =H_0 -H_4^{(b)}$, leading to the model in which magnon-magnon
interactions specific to  the model with a WZ term are omitted
(see \cite{PaperAnnPhys} for details).
The  one-loop self energy in this approximation can be written as
%------------------------------------------------------------------------------------------------------------------------
\bea
\widetilde{\Sigma}_{\text{RPA}}(\bm k) & = & 
\begin{array}{l} \vspace{0.95cm}
 {\includegraphics[scale=0.8]{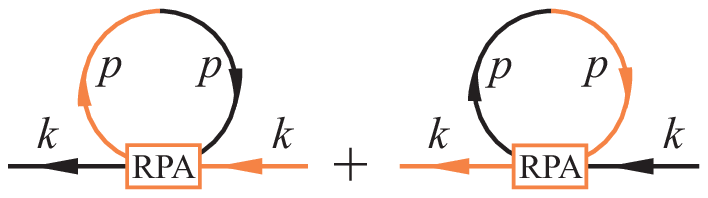}}
\end{array} \nonumber \\
& = & \frac{1}{S} \frac{v_0}{2 m_0} \int_{\bm p}
\Lan n_{\bm p} \Ran_0 \; \widehat{\bm p - \bm k}^2 \label{SigmaRPA},
\eea
%------------------------------------------------------------------------------------------------------------------------
where $\begin{array}{r}
 {\includegraphics[scale=1.1]{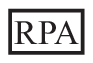}}
\end{array} = - \begin{array}{r}
 {\includegraphics[scale=1.1]{Diag_2.eps}}
\end{array}$ denotes the corresponding vertex.
Of course, this cannot be the final result, since (\ref{SigmaRPA})
implies the gap in magnon spectrum. To get physically 
acceptable self-energy, we must impose the constrain
$ \TI := v_0 \int_{\bm p}\Lan n_{\bm p} \Ran_0  \widehat{\bm p}^2 = 0$. But this 
is equivalent to $v_0 \int_{\bm p}\Lan n_{\bm p} \Ran_0 \gamma(\bm p) = 
v_0 \int_{\bm p}\Lan n_{\bm p} \Ran_0 \equiv \Lan n_{\bm x} \Ran$, 
and the constrain actually introduces site-independent mean
number of magnons, resembling the  replacement of 
the operator $S_{\bm x}^z(t)$ with site-independent average
$\Lan S^z \Ran$ in TGF theory.
Renormalized magnon energies become
%------------------------------------------------------------------------------------------------------------------------
\bea
\omega_{\text{RPA}}(\bm k) = \omega_0 (\bm k) - \Sigma_{\text{RPA}}(\bm k) 
= \frac{\widehat{\bm k}^2}{2 m_0} \frac{S-\Lan n_{\bm x} \Ran}{S} 
\label{OmegaRPA},
\eea
%------------------------------------------------------------------------------------------------------------------------
which is indeed the low-temperature limit of RPA result.
If we reverse the steps leading us to the effective
Hamiltonian of lattice magnon fields, we see that RPA can be taught of
as arising from the effective Lagrangian
%-------------------------------------------------------------------------------------------------------------------
\bea
\L_{\text{eff}}^{\text{RPA}} & = & \frac \Sigma 2\left(\partial_t U^1 U^2 - \partial_t U^2 U^1\right)
-\frac{F^2}{2} \partial_{\alpha} U^i \partial_{\alpha} U^i \nonumber \\
& - & \frac{F^2}{4} \bm \pi^2(\bm x)  \Delta \bm \pi^2(\bm x). \label{LagrRPA}
\eea
%-------------------------------------------------------------------------------------------------------------------
which manifestly violates internal O(3) symmetry of the
Heisenberg spin Hamiltonian starting at
$\O(\bm p^2)$. 
We stress that the reduction of magnon-magnon interactions
is is crucial for thermodynamics of O(3) ferromagnets,
for spurious terms in low-temperature series for free energy are present
whether or not the constrain 
$v_0 \int_{\bm p}\Lan n_{\bm p} \Ran_0  \widehat{\bm p}^2 = 0$
is implemented \cite{PaperAnnPhys}. 

As noted in the Introduction, RPA is widely used tool for 
studying  magnetic systems 
\cite{Kunc, PRB,PRB2,EPJB,SSCCV,SSCTNRPA, PRBNemci,PRBGFIhle1,PRBGFIhle2,PRBGFIhle3,
PRBGFLowD1,PRBGFLowD2,PRBGFLowD3,PRBGFLowD4,PRBGFLowD5,PRBGYabl,
J1J2SSC,Plakida2014EPJB,EPJB,PRBGFLongRange,Devlin,PRB,EPJBGFAnis,
RMPMagSemi, PRBMagSemi, PRBnucl, PRBMagnEl1,PRBMagnEl2,PRB_itin1,Phil_Mag,Semi_2015}. 
Thus, to be able to correctly interpret
results obtained from RPA,
it is of interest to fully grasp
the approximation in itself.
The spin operator-based mathematical machinery of two-time temperature
Green's functions makes the physics involved less obvious, and this can 
lead to a wrong conclusion. For instance, the authors of a  recent paper \cite{LosePRB}
characterize RPA as a theory which "which
does not take account of spin correlations or magnon-magnon interactions".
This is at best misleading, 
since we have shown that
RPA corresponds to an effective field theory in which magnon-magnon
interactions   arising in the WZ term are omitted, but those
induced by the unimodular constraint are kept. Thus, the difference
between LSW and RPA comes from magnon-magnon interactions
described by $-H_4^{(b)}$ of  (\ref{HamEffLatt1}). The thermodynamic averages
in RPA are actually calculated using $H_{\text{RPA}} = H_0 -H_4^{(b)}$ and not
the full Heisenberg Hamiltonian, as one may  expect  
from the mathematical formalism of TGF theory \cite{Plakida2013,Kunc}.

%-------------------------------------------------------------------------------------------------------------------
\begin{figure*}
\bc 
\includegraphics[scale=0.95]{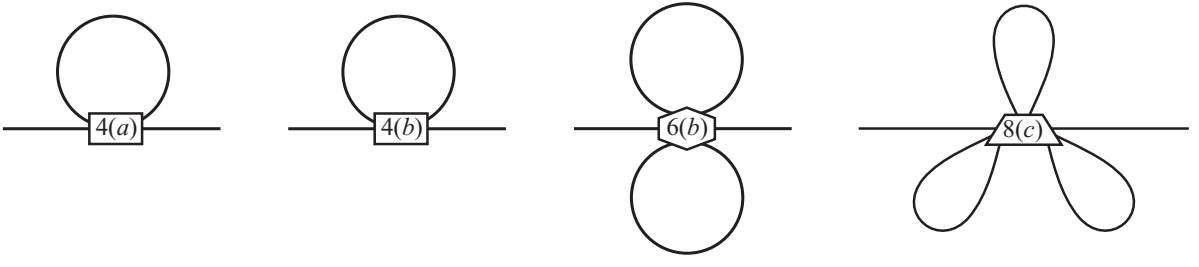}
\caption
{\label{fig2} The diagrams constituting magnon self-energy in the Callen approximation.}
\ec
\end{figure*}
%-------------------------------------------------------------------------------------------------------------------

%===============================================================================================================================================
\subsection{The Callen Approximation}\label{SecCA}
%===============================================================================================================================================

\subsubsection{Magnon Spectrum} \label{MagSpecCA}

Next we consider the Callen approximation (CA)
(See \cite{Kunc} and references therein),  which may also be
viewed as a linearization scheme
for EOM of spin operators.
The Callen's linearization approximately accounts for the short-range
fluctuations of the order parameter, apparently  neglected in RPA. 
As a result of CA, magnons acquire renormalized mass
%-------------------------------------------------------------------------------------------------------------------
\bea
m_{\text{CA}}^{-1} = m_{\text{RPA}}^{-1} \left[1+ 2 \frac{\Lan S^z \Ran}{2 S^2} \; v_0 \int_{\bm q}
\Lan n_{\bm q} \Ran_{\text{CA}} \gamma(\bm q) \right]
\label{mCA},
\eea 
%-------------------------------------------------------------------------------------------------------------------
with unchanged geometry of dispersion relation
%-------------------------------------------------------------------------------------------------------------------
\bea
\omega_{\text{CA}}(\bm k) = \frac{\widehat{\bm k}^2}{2 m_{\text{CA}}}. \label{OmegaCA}
\eea 
%-------------------------------------------------------------------------------------------------------------------
$\Lan S^z \Ran /{2 S^2}$, usually denoted as $\alpha(T)$, is a phenomenological
parameter introduced by Callen to get a "correction to RPA" \cite{Kalen63}
($\alpha(T)=0$ directly leads to RPA). Since the discussion from previous section
pointed out on the neglection of magnon-magnon interactions generated
by WZ term as a defining feature of RPA, the question of 
magnon-magnon interactions induced by CA naturally arises.
Now we examine it in detail.

First, we observe that the entire effect of Callen's linearization
again reduces to the pure renormalization of magnon mass. This
means that the magnon-magnon interactions are described solely
by one-vertex interactions (perturbations), i.e. sunset and other similar diagrams
are absent as in the case of RPA. Further, using 
$\Lan S^z \Ran \approx S- \Lan n_{\bm x} \Ran$ which is
certainly good approximation at low temperatures, we see
that $m_{\text{CA}}^{-1}$ contains terms linear, quadratic and
cubic in $\Lan n_{\bm q} \Ran$.
According to the  general structure of  one-vertex corrections
to the self-energy outlined in Sec. \ref{SecLMH}, 
the effective Hamiltonian that ought to produce
(\ref{OmegaCA}) through perturbation theory must contain terms
with six magnon operators. However, they are absent from 
(\ref{HamEffLatt0}), since products of  six magnon operators originating
from term with spatial derivatives     and WZ term precisely cancel.
This  is a special characteristic of the effective
Lagrangian (\ref{EffLagr}), and terms with six Goldstone fields 
generally appear in Lorentz-invariant theories \cite{Gerber,Smilga}.
The most general effective Hamiltonian with six magnon field operator and 
single lattice derivative  is
%-------------------------------------------------------------------------------------------------------------------
\bea
H_6 = H_6^{(a)} + H_6^{(b)} + H_6^{(c)}
\eea 
%-------------------------------------------------------------------------------------------------------------------
with
%-------------------------------------------------------------------------------------------------------------------
\bea
H_6^{(a)} & = & A_6 F^2 v_0 \sum_{\bm x} \left[ \bm \pi^2(\bm x) \right]^2 
\bm \pi (\bm x) \cdot \nabla^2 \bm \pi (\bm x), \nonumber \\ 
H_6^{(b)} & = & B_6 F^2 v_0 \sum_{\bm x} \left[ \bm \pi^2(\bm x) \right]^2 \nabla^2 \bm \pi^2(\bm x), \label{H_6} \\
H_6^{(c)} & = & C_6 F^2 v_0 \sum_{\bm x} \bm \pi^2(\bm x)\; \bm \pi (\bm x) \cdot \nabla^2
\left[\bm \pi(\bm x) \; \bm \pi^2(\bm x) \right], \nonumber
\eea 
%-------------------------------------------------------------------------------------------------------------------
where $A_6,B_6$ and $C_6$ are arbitrary constants. We can safely choose $C_6=0$,
since this kind of interaction is not generated neither by $\partial_\alpha \bm U
\cdot \partial_\alpha \bm U$, nor by WZ term of (\ref{EffLagr}).
Also, we may set $A_6=0$  as the the contribution of $H_6^{(a)}$ to
the self-energy
%-------------------------------------------------------------------------------------------------------------------
\bea
\Sigma_6^{(a)}(\bm k) & = & \begin{array}{l} \vspace{-0.15cm}
 {\includegraphics[scale=0.8]{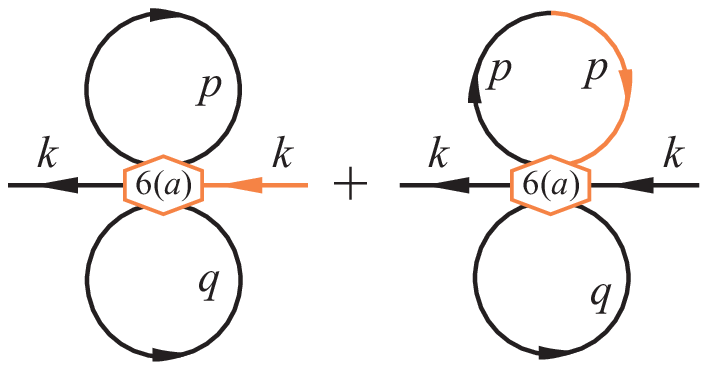}}
\end{array} \label{H6CA(a)}  \\
& = & \frac{48 A \Lan n_{\bm x} \Ran ^2}{S^2} \; \frac{\widehat{\bm k}^2}{2 m_0} \nonumber
 +  \frac{96 A \Lan n_{\bm x} \Ran }{S^2} \; \frac{ v_0}{2 m_0}
\int_{\bm q} \Lan n_{\bm q} \Ran_0 \widehat{\bm q}^2
\eea 
%-------------------------------------------------------------------------------------------------------------------
contains the term $\propto \widehat{\bm k}^2 \Lan n_{\bm x} \Ran ^2 S^{-2}$,
which does not appear in $\omega_{\text{CA}}(\bm k)$. As in earlier sections,
$\begin{array}{r}
 {\includegraphics[scale=1.1]{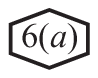}}
\end{array}$    denotes  the corresponding vertex.

As far as eight-magnon terms are concerned, one  easily verifies that
$H_8^{(a)}$ and $H_8^{(b)}$ can  not produce desired terms
in magnon self-energy. The corrections arising from $H_8^{(b)}$
are listed in (\ref{Sigma8b-1})-(\ref{Sigma8b-3}), revealing
redundant contribution from (\ref{Sigma8b-3}). Also, $H_8^{(a)}$
gives unacceptable term $\propto \widehat{\bm k}^2 \Lan n_{\bm x} \Ran^3 S^{-3} $.
In fact, the only way to reproduce $\omega_{\text{CA}}(\bm k)$
through the perturbation theory is to include the new eight-magnon
term 
%-------------------------------------------------------------------------------------------------------------------
\bea
H_8^{(c)} & = & C_8 F^2 v_0 \sum_{\bm x} \left[ \bm \pi^2(\bm x) \right]^3 
 \nabla^2 \bm \pi ^2(\bm x) \label{H_8(C)}.
\eea 
%-------------------------------------------------------------------------------------------------------------------
If the remaining coefficients from (\ref{H_6}) and (\ref{H_8(C)}) are chosen to 
be $B_6=1/32$ and $C_8=-1/576$, the  Hamiltonian
%-------------------------------------------------------------------------------------------------------------------
\bea
H_{\text{CA}} = H_0+  H_4^{(a)} + H_4^{(b)} + H_6^{(b)} + H_8^{(c)}, \label{H8CA}
\eea 
%-------------------------------------------------------------------------------------------------------------------
with suitable constrain on self-energy, yields $\omega_{\text{CA}}(\bm k)$
 through the perturbative corrections.
To confirm this, we simply evaluate the one-vertex self-energy for
diagrams depicted at Fig. \ref{fig2}.
Since
\begin{widetext}
%------------------------------------------------------------------------------------------------------------------------
\bea
\begin{array}{l} \vspace{0.95cm}
 {\includegraphics[scale=0.8]{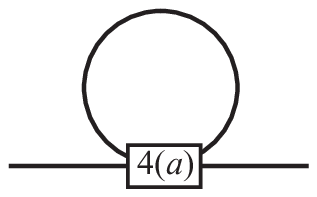}} \end{array} & = &
\begin{array}{l} \vspace{0.95cm}
 {\includegraphics[scale=0.8]{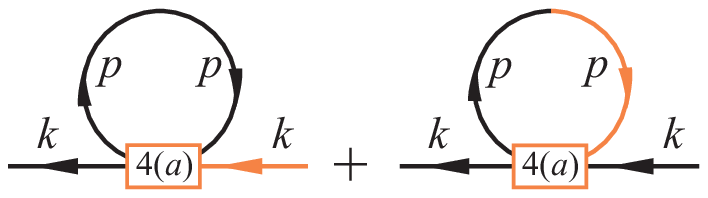}} \end{array} =
\frac{1}{S} \frac{v_0}{2 m_0} 
\int_{\bm q} \Lan n_{\bm q} \Ran_0 \; 
\left[\widehat{\bm k}^2 + \widehat{\bm q}^2 \right], \\
%\eea 
%-------------------------------------------------------------------------------------------------------------------
%------------------------------------------------------------------------------------------------------------------------
%\bea
\begin{array}{l} \vspace{0.95cm}
 {\includegraphics[scale=0.8]{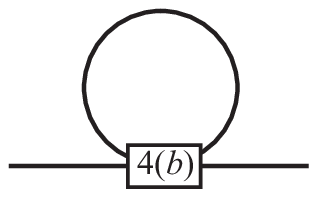}} \end{array} & = & 
\begin{array}{l} \vspace{0.95cm}
 {\includegraphics[scale=0.8]{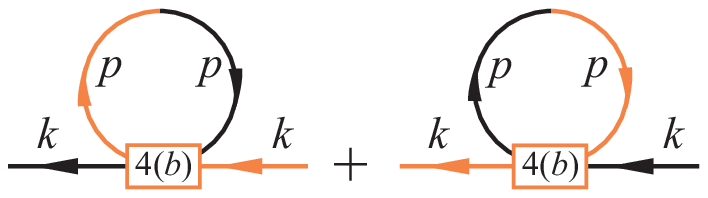}} \end{array}  =  -
\frac{1}{S} \frac{v_0}{2 m_0} 
v_0 \int_{\bm p} \Lan n_{\bm p} \Ran_0 \; \widehat{\bm k - \bm p}^2 ,%\\
\eea 
%-------------------------------------------------------------------------------------------------------------------
%------------------------------------------------------------------------------------------------------------------------
\bea
\begin{array}{l} \vspace{-0.15cm}
 {\includegraphics[scale=0.8]{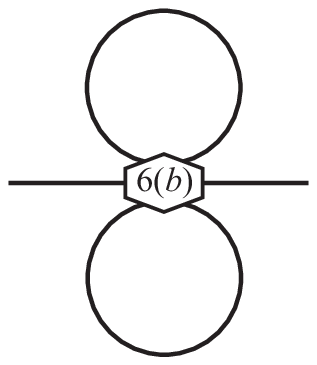}} \end{array} & = &
 \begin{array}{l} \vspace{-0.15cm}
{\includegraphics[scale=0.8]{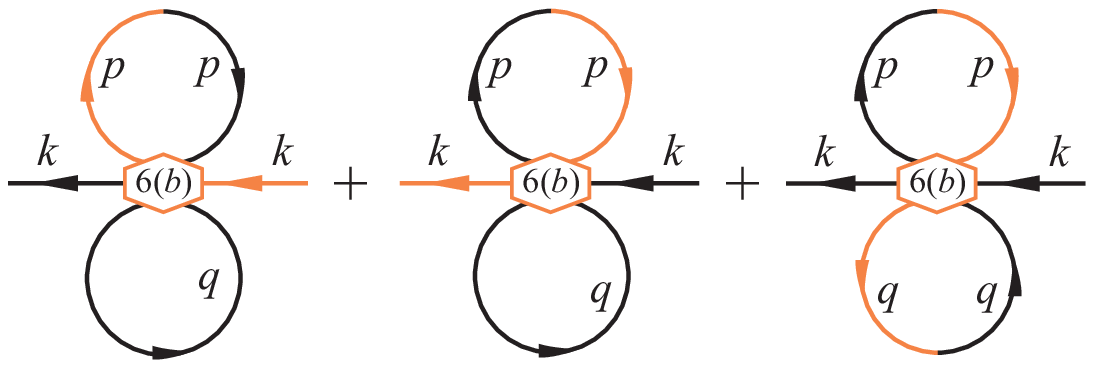}} \end{array} \nonumber \\
& = & \frac{1}{S^2}\;\frac{2 \Lan n_{\bm x} \Ran_0}{2 m_0}\; v_0 \int_{\bm p}
\Lan n_{\bm p} \Ran_0 \; \widehat{\bm k - \bm p}^2 
+ \frac{1}{S^2}\; \frac{1}{2 m_0} \; v_0^2 \int_{\bm p, \bm q}
\Lan n_{\bm p} \Ran_0 \Lan n_{\bm q} \Ran_0 \; \widehat{\bm p - \bm q}^2,    \\
\begin{array}{l} \vspace{0.07cm}
{\includegraphics[scale=0.8]{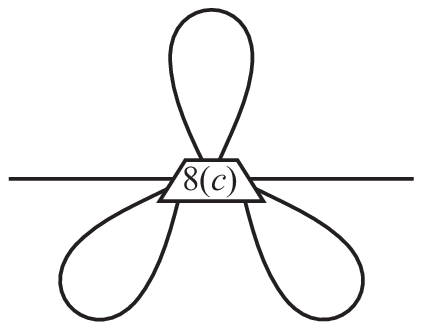}} \end{array} & = &
\begin{array}{l} \vspace{0.07cm}
{\includegraphics[scale=0.8]{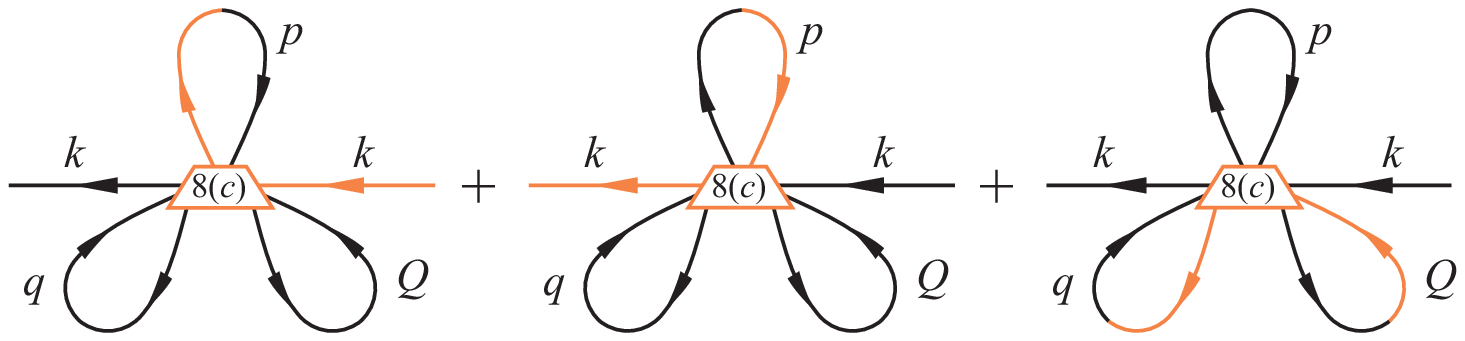}} \end{array} \nonumber \\
& = & -\frac{1}{S^3}\;\frac{\Lan n_{\bm x} \Ran_0^2}{2 m_0}\; v_0 \int_{\bm p}
\Lan n_{\bm p} \Ran_0 \; \widehat{\bm k - \bm p}^2 
- \frac{1}{S^3}\; \frac{ \Lan n_{\bm x} \Ran  }{2 m_0} \; v_0^2  \int_{\bm q, \bm Q}
\Lan n_{\bm q} \Ran_0 \Lan n_{\bm Q} \Ran_0 \; \widehat{\bm q - \bm Q}^2  ,
\eea
%------------------------------------------------------------------------------------------------------------------------
by putting all contributions together, we find:
%------------------------------------------------------------------------------------------------------------------------
\bea
\widetilde{\Sigma}_{\text{CA}}(\bm k) & = & \frac{1}{S} \frac{\widehat{\bm k}^2}{2 m_0}
\frac{|\lambda|^2}{2 D} v_0 \int_{\bm q} \Lan n_{\bm q} \Ran_0 \widehat{\bm q }^2 
 +  \frac{\Lan n_{\bm x} \Ran}{S^2}  \frac{\widehat{\bm k}^2}{2 m_0}
v_0 \int_{\bm q} \Lan n_{\bm q} \Ran_0 \gamma(\bm q)
\left[  2- \frac{\Lan n_{\bm x} \Ran}{S} \right] \nonumber \\
& + & \frac{1}{S^2} 
\left[4 \Lan n_{\bm x} \Ran-  \frac{|\lambda|^2}{2 D} \TI   \right]
 \frac{\TI}{2 m_0}
-\frac{\Lan n_{\bm x} \Ran}{S^3} 
\left[3 \Lan n_{\bm x} \Ran-  \frac{|\lambda|^2}{2 D} \TI  \right]
 \frac{\TI}{2 m_0} \label{TildeCA}, 
 \eea
%------------------------------------------------------------------------------------------------------------------------
\end{widetext}
where, as in the case of RPA, 
$\TI  =  v_0 \int_{\bm q} \Lan n_{\bm q} \Ran_0 \widehat{\bm q}^2$.
The absence of external momentum in  terms proportional
to $\TI$ and $\TI^2$ makes the self-energy
(\ref{TildeCA}) unphysical. 
These arise from two and three loop corrections 
that don't include true  interactions generated by WZ term and
unimodular constrain thus 
explicitly breaking O(3) symmetry at 
six and eight magnon field operators (see equations
(\ref{EffLagrBold2})-(\ref{EffLagrBold6}) below).
Therefore,  the contributions
to the  self energy calculated with $H_6^{(b)}$ and $H_8^{(c)}$
must be supplemented with additional constrain $\TI = 0$
for a magnon spectrum to display the Goldstone mode.
With restrictions on $\TI$, the self-energy (\ref{TildeCA}) reduces to
%------------------------------------------------------------------------------------------------------------------------
\bea
\hspace*{-0.4cm}\Sigma_{\text{CA}}(\bm k) & = &\frac{\widehat{\bm k}^2}{2 m_0}
\left[ \frac{\Lan  n_{\bm x} \Ran}{S} - \frac \I S 
+  \frac{2 \Lan  n_{\bm x} \Ran \I}{S^2}
-\frac{\I \Lan  n_{\bm x} \Ran^2}{S^3} \right],  \label{SigmaCA}
 \eea
%------------------------------------------------------------------------------------------------------------------------
where $\I   =  v_0 \int_{\bm q} \Lan n_{\bm q} \Ran_0 \gamma_D(\bm q)$.
The magnon energies in this approximation read
%------------------------------------------------------------------------------------------------------------------------
\bea
\omega_{\text{CA}}(\bm k) & = & \frac{\widehat{\bm k}^2}{2 m_0} - \Sigma_{\text{CA}}(\bm k) \nonumber \\
& = & \frac{\widehat{\bm k}^2}{2 m_0} \frac{S-\Lan  n_{\bm x} \Ran}{S}
\left[ 1+   \frac{S-\Lan  n_{\bm x} \Ran}{S^2} \I   \right] \label{OmegaCAlowT}
 \eea
%------------------------------------------------------------------------------------------------------------------------
and we recognize (\ref{OmegaCAlowT}) as low-energy limit
of Callen's result (\ref{OmegaCA}).
From the  point of view of an effective theory, 
all terms in magnon self energy arising from
symmetry-violating parts of the magnon Hamiltonian
must be constrained with $\TI = 0$ to mimic
O(3)-invariant perturbation theory.
It is well known that fine tunings of 
this sort are a necessity if the approximation in question is
too crude to encapsulate fundamental properties 
of a model under study \cite{Wen}.
In the present case it is  the replacement  of the lattice magnon 
Hamiltonian (\ref{HamEffLatt0}) with (\ref{H8CA}) that destroys spin-rotational invariance
and affects the magnon self-energy (\ref{TildeCA}).
Note, however, that the constrain $\TI=0$ should not be enforced
on leading term in (\ref{TildeCA}), since $H_4^{(a)}+H_4^{(b)}$
preserves spin-rotational invariance at four magnon operators.
In contrast to RPA, the constrain $\TI = 0$ merely eliminates
the gap and should not be interpreted as introducing the site-independent number
of magnons.

Thus, Callen's modification of RPA actually consists in adding
 three more interacting terms to $H_{\text{RPA}}$. The one
with four magnon fields ($H_4^{(a)}$) carries interactions 
generated by the WZ term and preserves spin-rotational
invariance up to one-loop corrections to the self energy. On
the other hand, $H_6^{(b)}$ and $H_8^{(c)}$
describe additional interactions which are, however, 
not generated by the effective Lagrangian (\ref{EffLagr}),
as seen from (\ref{HamEffLatt1})-(\ref{HamEffLatt2}).
The differences between
$\Sigma_{\text{CA}}(\bm k)$, $\Sigma_{\text{RPA}}(\bm k)$,
and two-loop result (\ref{2LoopSEM}),
which are obvious, are thus fully explained in terms 
of magnon-magnon interactions.

\subsubsection{Low-Temperature Thermodynamics} \label{LowTCA}

To gain further insight into the CA, we shall consider the low-temperature
series for free energy comparing, where necessary, the prediction of CA with those
of true magnon interaction theory\footnote{The accompanying analysis of 
low-temperature thermodynamics in the RPA model, 
defined in Sec. \ref{SecRPA}, can be found
in \cite{PaperAnnPhys}}. Starting from the  magnon Hamiltonian (\ref{HamEffLatt0})
on simple cubic lattice
we obtain Hamiltonian density \cite{PaperAnnPhys,DombreRead} up to $\bm p^6$   
%-------------------------------------------------------------------------------------------------------------------
\bea
\H_0 & = & \frac{F^2}{2} \partial_\alpha \bm \pi \cdot \partial_\alpha \bm \pi
- \Sigma \mu H \left( 1- \bm \pi^2/2  \right), \nonumber
\\ 
\H^{[2]} & = & \frac{F^2}{8} \left[ \bm \pi^2 \bm \pi \cdot \Delta \bm \pi
- \bm \pi^2 \Delta \bm \pi^2  \right] \nonumber \\
& + & \frac{F^2}{32} \left[ \bm \pi^4  \Delta \bm \pi^2
- \frac{1}{18} \bm \pi^6 \Delta \bm \pi^2  \right] \nonumber \\
\H^{[4]}  & = & -l_1 \partial_\alpha^2 \bm \pi \cdot \partial_\alpha^2 \bm \pi \label{HContP6} \\
& + & 
 \frac{l_1}{4} \left[ \partial_\alpha^2 \left( \bm \pi^2 \bm \pi \right)
\cdot \partial_\alpha^2 \bm \pi
- \partial_\alpha^2 \bm \pi^2 \partial_\alpha^2 \bm \pi^2 \right] \nonumber \\
&+& \frac{l_1}{16} \left[ \partial_\alpha^2  \bm \pi^4  \partial_\alpha^2 \bm \pi^2
- \frac{1}{18} \partial_\alpha^2 \bm \pi^6 \partial_\alpha^2 \bm \pi^2 \right] \nonumber  \\
\H^{[6]}  & = &  c_1 \partial_\alpha^3 \bm \pi \cdot \partial_\alpha^3 \bm \pi \nonumber \\ 
& + &  \frac{c_1}{4} \left[- \partial_\alpha^3 \left( \bm \pi^2 \bm \pi \right)
\cdot \partial_\alpha^3 \bm \pi
+ \partial_\alpha^3 \bm \pi^2 \partial_\alpha^3 \bm \pi^2 \right] \nonumber \\ 
&+& \frac{c_1}{16} \left[- \partial_\alpha^3  \bm \pi^4  \partial_\alpha^3 \bm \pi^2
+ \frac{1}{18} \partial_\alpha^3 \bm \pi^6 \partial_\alpha^3 \bm \pi^2 \right], \nonumber
\eea
%-------------------------------------------------------------------------------------------------------------------
with external magnetic field along $z-$axis,
 formal values of coupling constants $l_1 = F^2 a^2/24$,  $c_1 = F^2 a^4/720$
and $\Delta=\partial_\alpha \partial_\alpha$.
Given the Hamiltonian density (\ref{HContP6}) and knowing that the loops are suppressed
 by $D=3$ powers of momentum (see (\ref{Prop})), we are in the 
position to set up a diagrammatic series for the free energy density
$f=F/V$ \cite{Kapusta,PaperAnnPhys}. The magnon vertices in
continuum field-theoretic setting will be marked by
corresponding power of momentum with exception of those 
generated by $\H^{[2]}$, denoted by a dot.

The diagrams split into  two categories \cite{PaperAnnPhys}. 
The first one collects all one-loop diagrams with two-magnon vertices which
describe lattice anisotropies.
It is easy to see that their contribution to the free energy of
O(3) HFM on a simple cubic lattice yields terms whose
powers of temperature are $5/2, 7/2, 9/2$ and so on.
Moreover, the coefficients of two-magnon terms are the same in CA 
as in true magnon-interaction theory (compare (\ref{HContP6})
with equation (71) of \cite{PaperAnnPhys}).
It is therefore not surprising that CA exactly reproduces first 
three terms in low-temperature series for spontaneous magnetization
\cite{Kalen63}, as they are a simple consequence of the
lattice anisotropies.

Much more  interesting are multiloop diagrams as they arise  due to magnon-magnon
interactions. 
Out of two-loop diagrams, 
the first one that describes the magnon-magnon interaction,
and is a potential candidate for $T^4$ term, comes from four-magnon 
parts of $\H^{[2]}$. Yet, it vanishes since contributions 
from unimodular constraint and WZ term precisely cancel
 as in true magnon interaction theory  \cite{PaperAnnPhys,Hofmann2,Hofmann3}. 
The fact that low-temperature expansions obtained by Callen \cite{Kalen63} do
contain $T^4$ term for $S = 1/2$
should be ascribed to the pure artifact of TGF formalism \footnote{Actually, Callen
found corresponding spurious $T^3$ term in low-temperature series for spontaneous
magnetization and not directly the $T^4$ term discussed here}.
Thus, the first nonzero contribution to the free energy arising
solely from magnon-magnon interactions in CA is the Dyson term
(proportional to $T^5$),
the two-loop diagram of $\H^{[4]}$ \cite{PaperAnnPhys}:
%-------------------------------------------------------------------------------------------------------------------
\bea
\delta f_5 & = & -T \begin{array}{l}
%\vspace{1.1cm} 
\includegraphics[scale=0.8]{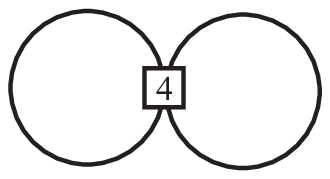} 
\end{array} \label{DysonCA}  \\
& = & - \frac{9 l_1  \pi^3}{2 \Sigma^2} \left(\frac{1}{2 \pi} \right)^{6}
 \left( \frac{\Sigma }{F^2} \right)^{5}  
 \left[ \sum_{n = 1}^{\infty} \frac{\e^{-\mu H n /T}}{n^{5/2}} \right]^2
  T^{5}.
\nonumber
\eea
%-------------------------------------------------------------------------------------------------------------------
The next two-loop diagram, including four-magnon vertex of $\H^{[6]}$ as well as two
and four magnon vertices of $\H^{[4]}$ gives $T^6$ term (see \cite{PaperAnnPhys}).
Just as $\delta f_5$, it occurs both in CA and in true magnon interaction theory.
Still, this is not the leading correction to the Dyson term.

The next term describing magnon-magnon interactions comes from
three-loop diagrams. In true magnon interaction theory, this is
the three loop diagram of $\H^{[2]}$ \cite{PaperAnnPhys,Hofmann3}
%-------------------------------------------------------------------------------------------------------------------
\bea
\delta f_{11/2} & = & -T \begin{array}{l}
%\vspace{1.1cm} 
\includegraphics[scale=0.7]{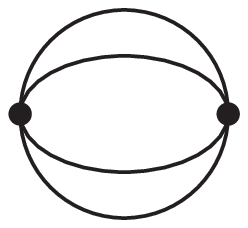} 
\end{array}  
 =  - \frac{1}{\Sigma^2} \left( \frac{\Sigma }{F^2} \right)^{9/2}
{I}(h)   T^{11/2}, \nonumber \\
h & = & \mu H/T, \label{BeyondDyson} 
\eea
%-------------------------------------------------------------------------------------------------------------------
where, in the absence of external magnetic field $H$, $I(h=0) = 5.3367 \times10^{-6}$
\cite{PaperAnnPhys}. As the analysis from previous section showed that
the CA magnon dispersion emerges through the perturbation theory only
if interactions described by two-vertex (and higher) diagrams are 
neglected \footnote{This does not exclude multi-vertex diagrams if all except
single vertex describe lattice anisotropies},
the CA low-temperature series for free energy does
not contain $T^{11/2}$ term of (\ref{BeyondDyson}).
However, a term proportional to $T^{11/2}$ is present in CA,
and it is generated by three-loop diagram of $\H^{[2]}$. Since
%-------------------------------------------------------------------------------------------------------------------
\bea
\begin{array}{l}
%\vspace{1.1cm} 
\includegraphics[scale=0.7]{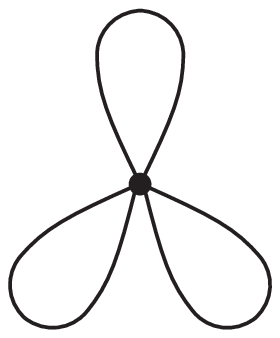} 
\end{array}  
 = \begin{array}{l}
%\vspace{1.1cm} 
\includegraphics[scale=0.7]{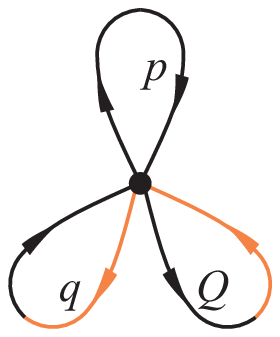} 
\end{array} +
\begin{array}{l}
%\vspace{1.1cm} 
\includegraphics[scale=0.7]{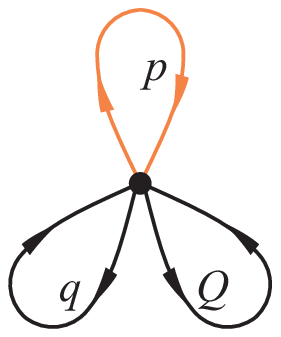} 
\end{array}  , \label{BeyondDysonCAdiag} 
\eea
%-------------------------------------------------------------------------------------------------------------------
with first colored contraction  appearing four times
and second one vanishing, we find
%-------------------------------------------------------------------------------------------------------------------
\bea
\delta f_{11/2}^{\text{CA}}& = &-\frac{2}{\Sigma^2} \left( \frac{\Sigma }{F^2} \right)^{9/2}
\frac{\pi^{9/2}}{\left(  2 \pi \right)^9} 
 \left[ \sum_{n = 1}^{\infty} \frac{\e^{-\mu H n /T}}{n^{3/2}} \right]^2 \nonumber \\
&\times &  \left[ \sum_{n = 1}^{\infty} \frac{\e^{-\mu H n /T}}{n^{5/2}} \right] T^{11/2}
\label{BeyondDysonCA} 
\eea
%-------------------------------------------------------------------------------------------------------------------
By comparing (\ref{BeyondDyson}) and (\ref{BeyondDysonCA}) for
$H=0$, we that CA overestimates 
the leading correction to  Dyson's term nearly by a factor of 39.
Obtaining expression (\ref{BeyondDysonCA}) within TGF formalism
is somewhat tedious, but rather straightforward 
within magnon perturbation theory, i.e. when appropriate
set of magnon-magnon interactions have been identified.

Finally, we note that Hamiltonian (\ref{HContP6}) can be obtained from
the Lagrangian
%-------------------------------------------------------------------------------------------------------------------
\bea
\L = \L^{[2]} + \L^{[4]} +\L^{[6]} \label{LContP6}
\eea
%-------------------------------------------------------------------------------------------------------------------
where
%-------------------------------------------------------------------------------------------------------------------
\bea
\L^{[2]} & = & \Sigma \frac{\partial_t U^1 U^2 - \partial_t U^2 U^1}{1+U^3}
-\frac{F^2}{2} \partial_{\alpha} \bm U \cdot \partial_{\alpha} \bm U \nonumber 
 + \Sigma \mu H U^3 \\
 & + & \frac{F^2}{32} \partial_{\alpha}  \bm \pi ^4 \partial_{\alpha}  \bm \pi ^2
-  \frac{F^2}{576} \partial_{\alpha}  \bm \pi ^6 \partial_{\alpha}  \bm \pi ^2 \label{EffLagrBold2}\\
& - & \frac{F^2}{128} \partial_{\alpha}  \bm \pi ^4 \partial_{\alpha}  \bm \pi ^4 
+ \frac{F^2}{128} \partial_{\alpha} 
\left( \bm \pi ^6  \bm \pi \right) \cdot \partial_{\alpha}  \bm \pi \nonumber, \\
\L^{[4]} & = & l_1 \partial_{\alpha}^2 \bm U \cdot \partial_{\alpha}^2  \bm U 
 -  \frac{l_1}{16} \left[ \partial_{\alpha}^2 \bm \pi^4 \partial_{\alpha}^2 \bm \pi^2
-\frac{1}{18} \partial_{\alpha}^2 \bm \pi^6 \partial_{\alpha}^2 \bm \pi^2  \right] \nonumber \\
& - & \frac{l_1}{64} \left[\partial_{\alpha}^2 \left( \bm \pi^6 \bm \pi  \right) \cdot
\partial_{\alpha}^2 \bm \pi - \partial_{\alpha}^2 \bm \pi^4 \partial_{\alpha}^2 \bm \pi^4    \right],
 \label{EffLagrBold4}  \\
\L^{[6]} & = & c_1  \bm U \cdot \partial_{\alpha}^3 \partial_{\alpha}^3  \bm U 
-  \frac{c_1}{16} \left[ \bm \pi^4 \partial_{\alpha}^3 \partial_{\alpha}^3 \bm \pi^2
-\frac{1}{18}  \bm \pi^6 \partial_{\alpha}^3 \partial_{\alpha}^3 \bm \pi^2  \right] \nonumber \\
& - & \frac{c_1}{64} \left[  \bm \pi^6 \bm \pi   \cdot
\partial_{\alpha}^3 \partial_{\alpha}^3 \bm \pi -  \bm \pi^4 \partial_{\alpha}^3 \partial_{\alpha}^2 \bm \pi^4   \right],
 \label{EffLagrBold6} 
\eea
%-------------------------------------------------------------------------------------------------------------------
It is seen that each $\L^{[2]}, \L^{[4]}$ and $\L^{[6]}$ contain terms which 
manifestly violate internal symmetry of the Heisenberg model, just
as in the case of RPA. 

The explicit expressions  from 
Sec. \ref{SecRPA} and Sec. \ref{SecCA} should supplement  imprecise statements  relating
RPA to CA
found in the literature and 
dating  as back as Callen's
original paper \cite{Kalen63}. For example, Tahir-Kheli
introduces CA as an approximation which takes "into
account the fluctuations of $S_{\bm n}^z$ around its
average $\Lan S^z \Ran$" \cite{Tahir-Kheli}. 
Further, it is claimed in a recent paper
\cite{LosePRB}  that "CA takes, to some extent, 
account of magnon-magnon interactions", without
actually specifying their type. 
Also, the point of view in standard and
up-to-date reference on TGF method \cite{Kunc} is that CA
"takes some higher-order correlations  into account".
By striping the issue of RPA/CA relationship down
to the problem of magnon-magnon interactions,
we have shown that  effective Hamiltonians (i.e. Lagrangians) that generate 
RPA and CA through perturbation theory violate
O(3) symmetry of the Heisenberg model at $\O(\bm p^2)$.
A difference is, of course, that manifest
symmetry breaking in CA takes place at six and eight magnon terms 
developing deviations from true magnon interaction theory
at three-loop corrections to the free energy, while 
RPA  breaks down already at two loop.

%===============================================================================================================================================
\subsection{Comparison with quantum Monte Carlo simulation}\label{QMCB}
%===============================================================================================================================================

We conclude the analysis in this section by comparing predictions
of effective field theories for RPA and CA with quantum Monte Carlo (QMC)
simulations and true magnon interaction theory for the O(3) ferromagnet
free energy. This will allow 
us to infer the influence of RPA and CA-type magnon-magnon interactions
on thermodynamics beyond leading order expansions
in $T$ discussed so far. The usefulness of the lattice EFT pursued
here will become clear when we look at results away from low-temperature
sector, i.e. when the lattice structure becomes resolved by magnons.
While  equations given bellow apply equally for arbitrary lattice, exchange
integral $J$ and localized spin $S$, 
all numerical calculations are conducted for the $S=1/2$ Heisenberg
ferromagnet on a simple cubic lattice with $J=1$. The parameters
$S=1/2$ and $J=1$, together with the lattice type,
define a concrete realization of the Heisenberg ferromagnet.
On the other hand, the parameters of EFT are $F$ and $\Sigma$
and  are related to those of original Heisenberg Hamiltonian by matching
[see the discussion bellow equation (\ref{HamEffLatt2})].

The free energy (per lattice site) corresponding to the non-interacting
theory is calculated with $H_0$ of (\ref{HamEffLatt0}). This is basically a
LSW result  and is given by
%-------------------------------------------------------------------------------------------------------------------
\bea
\frac{F_{\text{LSW}}}{N} = \frac{F_0}{N} + T v_0 \int_{\bm{k}}
\ln \left[ 1- \e^{-\omega_0(\bm k)/T} \right] \label{FLSW}.
\eea
%-------------------------------------------------------------------------------------------------------------------
To make comparison with QMC easier, we have normalized the non-interacting
 free energy by adding $F_0/N = -3/4$,
the ground state energy of $S=1/2$ and  $J=1$ Heisenberg ferromagnet on the simple cubic
lattice. We note that $F_{\text{LSW}}$ includes all one-loop
diagrams (in the sense of derivative expansion) permitted by the lattice symmetry.
 
The two-loop correction to $F_{\text{LSW}}$ in true magnon interaction theory (TMIT)
is found in \cite{PaperAnnPhys} to be
%-------------------------------------------------------------------------------------------------------------------
\bea
\frac{\delta F_{\text{TMIT}}}{N} = - \frac{1}{S} \frac{1}{2 m_0}
\frac{|\bm{\lambda}|^2}{4 D} \left[ v_0 \int_{\bm p}
\Lan n_{\bm p}
 \Ran_0\; \widehat{\bm p}^{\;2} \right]^2 \label{DeltaFTMIT}
\eea
%-------------------------------------------------------------------------------------------------------------------
and the two-loop correction corresponding to RPA is
%-------------------------------------------------------------------------------------------------------------------
\bea
\frac{\delta F_{\text{RPA}}}{N} & = & - T 
\begin{array}{l}
\vspace{0.0cm}\includegraphics[scale=0.8]{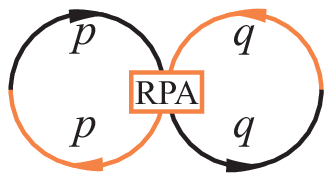}
\end{array} \nonumber \\
& = & -\frac{1}{2 S}\frac{v_0^2}{2 m_0}  \int_{\bm p, \bm q}
\Lan n_{\bm p} \Ran_0\;\Lan n_{\bm q}\Ran_0\; \widehat{\bm p - \bm q}^2
 \label{DeltaFRPA}
\eea
%-------------------------------------------------------------------------------------------------------------------
with RPA vertex defined bellow equation (\ref{SigmaRPA}). According to the 
discussion from Sec. \ref{SecCA}, CA type of magnon-magnon interactions
induce two, three and four-loop single vertex contributions to the ferromagnet free energy.
While the two-loop term is the same as in TMIT [equation (\ref{DeltaFTMIT})], 
the three and four loop corrections are given by
%-------------------------------------------------------------------------------------------------------------------
\bea
\frac{\delta F_{\text{CA}}^{3\text{loop}}}{N} & = & - T 
\begin{array}{l}
\vspace{0.0cm}\includegraphics[scale=0.8]{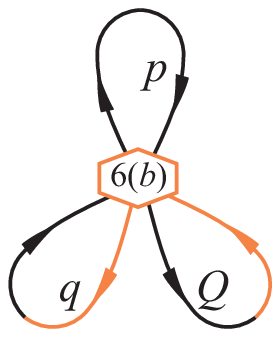}
\end{array} \nonumber \\
& = & -\frac{1}{S^2}\frac{\Lan n_{\bm x} \Ran }{2 m_0} v_0^2 \int_{\bm p, \bm q}
\Lan n_{\bm p} \Ran_0\;\Lan n_{\bm q}\Ran_0\; \widehat{\bm p - \bm q}^2
 \label{DeltaFCA3L}
\eea
%-------------------------------------------------------------------------------------------------------------------
and
%-------------------------------------------------------------------------------------------------------------------
\bea
\frac{\delta F_{\text{CA}}^{4\text{loop}}}{N} & = & - T 
\begin{array}{l}
\vspace{0.0cm}\includegraphics[scale=0.8]{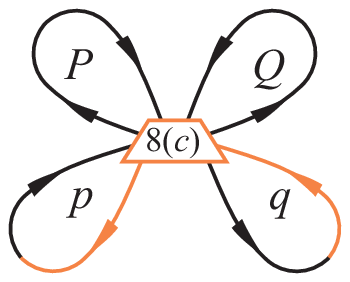}
\end{array} \nonumber \\
& = & \frac{1}{2S^3}\frac{\Lan n_{\bm x} \Ran^2 }{2 m_0} v_0^2 \int_{\bm p, \bm q}
\Lan n_{\bm p} \Ran_0\;\Lan n_{\bm q}\Ran_0\; \widehat{\bm p - \bm q}^2.
 \label{DeltaFCA4L}
\eea
%-------------------------------------------------------------------------------------------------------------------

Finally, the simulation was performed using 
stochastic series expansion within quantum Wang-Landau algorithm 
\cite{WangLandauPRL}  based on ALPS libraries \cite{ALPS_J_Stat_Mech}.
The cutoff is set to $5\times 10^4$, which, for the SC lattice with
$10^3$ sites, yields reliable results if $T \gtrsim 0.05  J$ (with $J=1$).
By letting the number of Wang-Landau  refinement steps   to be 18, 
the total number of sweeps becomes $\sim   10^{10}$ and the error bars
are much smaller than the line used to display the simulation results.
%-------------------------------------------------------------------------------------------------------------------
\begin{figure}
\bc 
\includegraphics[scale=0.75]{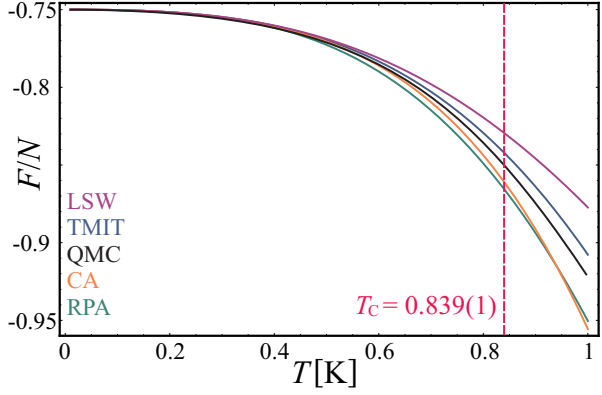}
\caption
{\label{fig3} The influence of magnon-magnon interactions on the free energy
of $S=1/2$ and $J = 1$ Heisenberg ferromagnet on the simple cubic lattice.
The purple (top) curve represents the non-interacting model (LSW), 
while blue and black one represent TMIT and QMC results. 
The orange and green lines correspond to the models with interactions
of CA and RPA type. The value of critical temperature $T_{\text{C}} = 0.839(1)$ K
is taken from \cite{WesselPRB}.}
\ec
\end{figure}
%-------------------------------------------------------------------------------------------------------------------

Results of  calculations are collected on Fig. \ref{fig3}. First, it is
obvious that all lattice magnon models (including LSW) develop only small deviations
from QMC almost up to $T=0.5$ K. This clearly confirms
our remarks about weakness of magnon-magnon interactions even in the
case of $S=1/2$ since the critical temperature  is, from QMC simulations, known to be 
$T_{\text{C}} \approx 0.84$ K  \cite{WesselPRB}. 
It is also evident that
TMIT produces results closest to QMC in wide temperature range, as expected,
demonstrating efficiency of the lattice magnon theory. 
Probably the most important observation concerning practical 
applications and interpretations of RPA and CA is that both approximations are based
on interacting theories which, thanking   the
loop structure of corrections and corresponding Hamiltonians, \emph{overestimate} the 
influence of magnon-magnon interactions [compare with equation (\ref{BeyondDysonCA})]. 
This is the main reason why
self-consistent descriptions of real compounds
based on RPA/CA, as a rule, predict higher values of the exchange integral $J$
than LSW or non-linear spin-wave theories based on boson
representations \footnote{In a self-consistent description of particular compound,
modeled by a Heisenberg Hamiltonian, all microscopic parameters such as the exchange integral $J$
are determined by matching with experimental data for some physical quantity. }.
For example, the non-linear spin wave theory in combination with
experimental data for 3D ferromagnet $\text{CrBr}_3$ predicts the intra- and interlayer
nearest neighbor exchange integrals
$J=8.25$ K and $J'=0.497$ K  \cite{PhysRev_CrBr3}. In contrast, self-consistently determined
exchange integrals within RPA are $J=12.38$ K and $J'=1.0$ K \cite{Irkin}.
Of course, the strength of magnon-magnon interactions in RPA and CA at 
higher temperatures is masked by the special form of self-consistent
equations for correlation functions.
Namely, RPA and CA solutions in TGF formalism
share many features with simple mean-field description at higher temperatures,
exemplified by the expressions for the  Curie temperature and the value of critical  index $\beta$.
As a result, both RPA and CA predict  somewhat higher value for the critical temperature
($T_{\text{C}}^{\text{RPA}} \approx 0.989$ K and $T_{\text{C}}^{\text{CA}} \approx 1.326$ K
for $J=1$ and $S=1/2$ ferromagnet on a simple cubic lattice).
Finally, we note that the similar conclusions concerning
low-temperature behavior of O(3) ferromagnet at low temperatures
were reached in \cite{EPL}.

%===============================================================================================================================================
\section{Perturbation theory for Type A magnons}\label{SecSOA}
%===============================================================================================================================================

Now we turn to the perturbation theory for type A
Goldstone bosons.  Basic diagrammatic rules remain the same,
but the definition of thermal propagator changes. Also,
we shall need to include more than one lattice Laplacian
per vertex as well as  couplings beyond nearest neighbors.
All this can be achieved within formalism of colored
propagators described in Sec. \ref{DiagDesc}.

\subsection{Kondo-Yamaji Equations}

The approximation of Kondo and Yamaji (KYA) represents an extension
of the EOM method in the sense the linearizations are conducted in
equations containing $\ddot{S}^{\pm}_{\bm n}$. The long-range order parameter
$\Lan S^z \Ran$ does not explicitly enter the magnon dispersion, making
this approach suitable for low-dimensional systems \cite{KYSOGF,JPSJSOGF}.
It is also a starting point in further developments
of TGF formalism \cite{PRBGFIhle1,PRBGFIhle2,PRBGFIhle3, JPSJSOGF}.
Here, the focus will be on one-dimensional ferromagnet. 
Strictly speaking, effective field theory is not directly applicable
to one-dimensional O(3) ferromagnet in the absence of external magnetic field,
since magnons acquire nonperturbatively generated gap.
However, a careful analysis from 
\cite{Hofmann6,Hofmann7,TakahashiPTPsupl,TakahashiPRL}
reveals that first few terms in low-temperature series for free energy are well defined even
in case of zero external field,  and KYA yields 
results in agreement with thermal Bethe-ansatz in this extreme low-$T$ sector \cite{Koma}.

The Kondo-Yamaji equations (KYE) for magnon dispersion
%-------------------------------------------------------------------------------------------------------------------
\bea
\omega_{\text{KY}}(k) & = & J \{ [1-\cos ka][1-\widetilde{c}_1+\widetilde{c}_2
-2 \widetilde{c}_1 \cos ka]   \}^{1/2}, \nonumber \\
\widetilde{c}_n & = & \widetilde{W} v_0 \int_{p} 
\frac{\coth{\frac{\omega_{\text{KY}}(p)}{2T}}}{2 \omega_{\text{KY}}(p)}
\gamma(n p)[1-\gamma(p)],
  \label{OmegaKJA} %\\
\eea
%-------------------------------------------------------------------------------------------------------------------
with $\widetilde{W} \approx  2 J$ and $\widetilde{c}_0  =  3/2$ at low temperatures, $\gamma(p)$ and $\int_p$ given
in (\ref{kSq}) and (\ref{Omega0}) for $D=1$,
 are derived from the assumption
of vanishing long range order (LRO) \cite{KYSOGF}. 
It turns out that is extremely useful to exploit the constrain $\widetilde c_0 = 3/2$
directly and to write KYA dispersion as
%-------------------------------------------------------------------------------------------------------------------
\bea
\omega_{\text{KY}}^2(k) & = & c^2 \widehat{k}^2 - c^2 \widehat{k}^2
\left[ 3\widetilde{c}_0 - \widetilde{c}_1 \right] \label{KYAGF}  \\
& + & \widehat{k}^2 c^2 \frac{a^6 \widetilde{W}v_0}{4}
\int_{p} \frac{\coth{\frac{\omega_{\text{KY}}(p)}{2T}}}{2 \omega_{\text{KY}}(p)}
\left[\widehat{p}^2 \right]^3  +
\widehat{k}^2 \widehat{k}^2 a^2 c^2 \widetilde{c}_1,  \nonumber \\
c^2 & = & \frac{J^2 a^2}{2} \nonumber.
\eea
%-------------------------------------------------------------------------------------------------------------------
We shall  demonstrate bellow
that KYE (\ref{OmegaKJA}) i.e. (\ref{KYAGF}), can be obtained from the effective Lagrangian description
of a ferromagnet chain if $\Sigma = 0$ is set in (\ref{EffLagr}).
When $\Sigma = 0$ the effective Lagrangian must include a term
with two temporal derivatives and it takes pseudo-Lorentzian form
%-------------------------------------------------------------------------------------------------------------------
\bea
\L_{\text{eff}} & = & \frac{F^2}{2} \partial_{t} U^i \partial_{t} U^i
-\frac{\gamma^2}{2} \partial_{x} U^i \partial_{x} U^i , \label{EffLagrLor}
%& + & \Sigma \mu H U^3. \nonumber
\eea
%-------------------------------------------------------------------------------------------------------------------
where, as before, $\bm U$ denotes the unit vector field and $F^2$
and $\gamma^2$ are constants  satisfying $c^2 = \gamma^2/F^2$.
In contemporary nomenclature, this Lagrangian describes
the dynamics of type A Goldstone bosons \cite{JapanciPRL,JapanciPRX}.

\subsection{Hamiltonian and perturbation theory for NN lattice model}

To develop the perturbation theory that yields KYE, we
construct the lattice magnon Hamiltonian. 
Since the Lagrangian (\ref{EffLagrLor}) contains $[\partial_t \bm U]^2$,
the Hamiltonian cannot be obtained   as in preceding sections.
Instead, we follow \cite{ChiralLoops} to get the interaction picture 
Hamiltonian. By putting it on a spatial lattice
(chain), we find the free part 
%-------------------------------------------------------------------------------------------------------------------
\bea
H_0 = \frac{v_0}{2 } \sum_{ x}\left[ \frac{1}{F^2} \Pi^a \Pi^a -
\gamma^2 \pi^a \nabla^2 \pi^a   \right] \label{H0Lor}
\eea
%-------------------------------------------------------------------------------------------------------------------
with (thermal) propagator
%-------------------------------------------------------------------------------------------------------------------
\bea
\hspace*{-0.2cm}D_{ab}(x - y, \tau_x - \tau_y) & \equiv & 
\Lan \mbox{T} \left\{ \pi^a( x,\tau_x) \pi^b (y,\tau_y) 
 \right\} \Ran_0  \label{PropLor} \\
& = &  \frac{\delta_{ab}}{\beta} \sum_{n = -\infty}^\infty 
\int_{q} \frac{\e^{\i q  ( x -  y) - \i
 \omega_n (\tau_x-\tau_y)}}{\omega^2_0(q) +  \omega_n^2}, \nonumber
\eea
%-------------------------------------------------------------------------------------------------------------------
free-magnon dispersion
%-------------------------------------------------------------------------------------------------------------------
\bea
\omega_0(q)  = c  \sqrt{\widehat{q}^2} \equiv c \widehat{q},  \label{Omega0Lor}
\eea
%-------------------------------------------------------------------------------------------------------------------
and nearest-neighbour (NN) interaction piece
%-------------------------------------------------------------------------------------------------------------------
\bea
\hspace*{-0.4cm} H_{\text{int}} = -\frac{\gamma^2}{8} v_0 \sum_{x} \bm \pi ^2 \nabla^2 \bm \pi^2
-\frac{F^2}{2}v_0 \sum_{ x} \left[ \bm \pi \cdot  \partial_\tau \bm \pi  \right]^2 \label{HintLor}
\eea
%-------------------------------------------------------------------------------------------------------------------
where  $\pi^a$ denotes the interaction picture
magnon field and  $\Pi^a$ is the corresponding canonical momentum.

To calculate one-loop self energy for interaction (\ref{HintLor})
we use diagrammatic rules outlined in Sec. \ref{DiagDesc}.
While the action of $\nabla^2$ on propagator lines
merely introduces  a factor of $\widehat{q}\;^2$ for external
or loop momenta, propagators with temporal derivatives
must be handled with care \cite{ChiralLoops}.
We find
%-------------------------------------------------------------------------------------------------------------------
\bea
\hspace*{-0.2cm}D_{ab}^\tau & \equiv & 
\Lan \mbox{T} \left\{ \partial_{\tau_x} \pi^a( x,\tau_x) \pi^b (y,\tau_y) 
 \right\} \Ran_0  \nonumber \\
& = &  \frac{\delta_{ab}}{\beta} \sum_{n = -\infty}^\infty 
\int_{q} D_\tau(q,\omega_n) \; \e^{\i q  ( x -  y) - \i
 \omega_n (\tau_x-\tau_y)}, \nonumber \\
\hspace*{-0.2cm}D_{ab}^{-\tau} & \equiv & 
\Lan \mbox{T} \left\{  \pi^a( x,\tau_x) \partial_{\tau_y} \pi^b (y,\tau_y) 
 \right\} \Ran_0   \label{PropLorTau} \\
& = &  \frac{\delta_{ab}}{\beta} \sum_{n = -\infty}^\infty 
\int_{q} D_{-\tau}(q,\omega_n) \; \e^{\i q  ( x -  y) - \i
 \omega_n (\tau_x-\tau_y)}, \nonumber \\
D_{ab}^{\tau  \tau} & \equiv & 
\Lan \mbox{T} \left\{ \partial_{\tau_x} \pi^a( x,\tau_x) \partial_{\tau_y} \pi^b (y,\tau_y) 
 \right\} \Ran_0  \nonumber \\
& = & \frac{\delta_{ab}}{\beta} \sum_{n = -\infty}^\infty 
\int_{q} D_{\tau \tau}(q,\omega_n) \; \e^{\i q  ( x -  y) - \i
 \omega_n (\tau_x-\tau_y)} \nonumber,
\eea
%-------------------------------------------------------------------------------------------------------------------
where
%-------------------------------------------------------------------------------------------------------------------
\bea
D_\tau(q,\omega_n) & = & - D_{-\tau}(q,\omega_n) =  
\frac{-\i \omega_n}{\omega^2_0(q) +  \omega_n^2} \nonumber \\
D_{\tau \tau}(q,\omega_n) & = & \frac{- \omega_0^2(q)}{\omega^2_0(q) +  \omega_n^2}
\label{DTauOmega}
\eea
%-------------------------------------------------------------------------------------------------------------------
Using propagators listed in (\ref{PropLorTau}) together with
(\ref{PropLor}) and action of lattice Laplacians, we obtain
one-loop self energy
%-------------------------------------------------------------------------------------------------------------------
\bea
\hspace*{-0.3cm}\widetilde{\Sigma}(k,\omega) & = & 
\frac{\gamma^2}{c^2} v_0
 \int_p \frac{\coth\frac{\omega_0(p)}{2 T}}{2 \omega_0(p)}
\left[ \omega_0^2(p) + \omega^2 - \omega_0^2(p-k) \right] \nonumber \\
& = & \gamma^2 v_0
 \int_p \frac{\coth\frac{\omega_0(p)}{2 T}}{2 \omega_0(p)}
\left[ \frac{a^2}{2} \widehat{k}^2  \widehat{p}^2 -\widehat{k}^2 + \frac{\omega^2}{c^2} \right]
\label{OneLoopSigmaNN}
\eea
%-------------------------------------------------------------------------------------------------------------------
and thus
%-------------------------------------------------------------------------------------------------------------------
\bea
\widetilde{\omega}^2(k) & = & \omega_0^2(k) - \widetilde{\Sigma}(k)  \label{TildeOmegaKYA} \\
& = & \widehat{k}^2 c^2 \left[ 1- F^2  v_0 
\int_p \frac{\coth\frac{\omega_0(p)}{2 T}}{2 \omega_0(p)} \frac{a^2}{2} \widehat{p}^2  \right] \nonumber
\eea
%-------------------------------------------------------------------------------------------------------------------
where $\widetilde{\Sigma}(k) \equiv \widetilde{\Sigma}(k,\omega_0(k))$ is the on-shell self
energy.
If we set $F^2 = 3 \widetilde{W} \approx 6 J$
and    interpret (\ref{TildeOmegaKYA}) as a
self-consistent equation for $\widetilde{\omega}(k)$,
%-------------------------------------------------------------------------------------------------------------------
\bea
\widetilde{\omega}^2(k) 
 & = &c^2 \widehat{k}^2 \left[ 1- F^2  v_0 
\int_p 
\frac{\coth\frac{\widetilde{\omega}(p)}{2 T}}{2 \widetilde{\omega}(p)}  \frac{a^2}{2}
\widehat{p}^2  \right] \nonumber \\
& = &  c^2 \widehat{k}^2 \left[1-3\widetilde{c_0}\right],
\eea
%-------------------------------------------------------------------------------------------------------------------
 we see that
the lattice model (\ref{HintLor}) with NN coupling accounts only for 
the leading term in $\Sigma_{\text{KY}}(k)$ listed in (\ref{KYAGF}).
Obviously, the full KYA self
energy can be obtained only with additional interactions in the one-loop
approximation.

\subsection{Perturbation theory for KYA model}

\subsubsection{Colored diagrams for NNN interaction}

Two additional conventions  will enable
us to imbed next-nearest neighbor (NNN)  interactions, 
as well as interactions containing more than
one lattice Laplacian into the framework
of colored diagrams. They  will also make diagrammatic rules 
suitable for potential higher-loop calculations. First, propagators
affected by $\nabla^2_{(2)}$ will be denoted by a double
colored line. Second, the presence of $\nabla^2_{(2)}$ or 
$\nabla^2$ will be represented by colored
edges of the pentagon, which denotes a vertex. For example,
typical interaction containing both $\nabla^2_{(2)}$ and
NN Laplacian $\nabla^2$ is $V = v_0 \sum_x
(\bm \pi \cdot  \nabla^2_{(2)} \bm \pi )
(\nabla^2 \bm \pi  \cdot \nabla^2 \bm \pi )$. Four distinct colored contractions
contribute to one-loop self energy generated by $V$.
These are
%-------------------------------------------------------------------------------------------------------------------
\bea
\begin{array}{l}
\vspace{1.3cm} 
\includegraphics[scale=0.5]{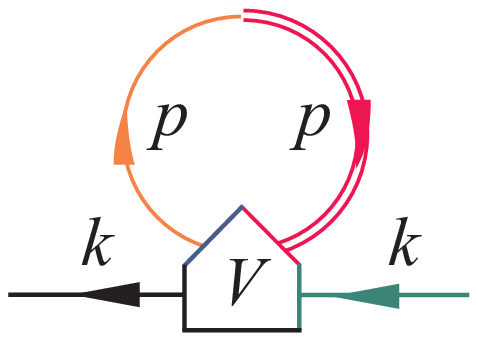} 
\end{array}  
& = & - 4 v_0 \int_p \frac{\coth\frac{\omega_{\text{0}}(p)}{2 T}}{2 \omega_{\text{0}}(p)}
\widehat{k}^2 \widehat{p}^2 \widehat{\; 2p \;}^2, \nonumber \\
\begin{array}{l}
\vspace{1.3cm} 
\includegraphics[scale=0.5]{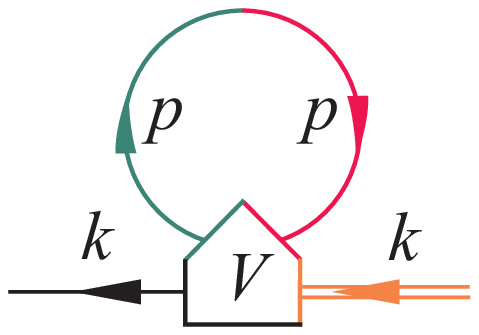} 
\end{array}  
& = &- 4 v_0 \int_p \frac{\coth\frac{\omega_{\text{0}}(p)}{2 T}}{2 \omega_{\text{0}}(p)}
\widehat{\; 2k \;}^2 \widehat{p}^2 \widehat{p}^2, \nonumber \\
\begin{array}{l}
\vspace{1.3cm} 
\includegraphics[scale=0.5]{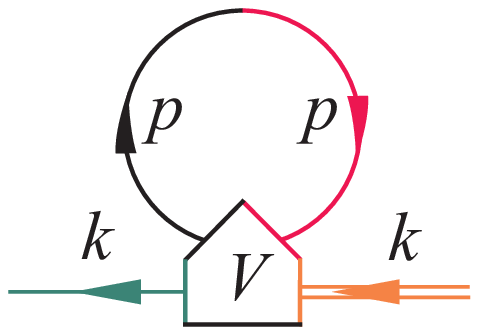} 
\end{array}  
& = &- 4 v_0 \int_p \frac{\coth\frac{\omega_{\text{0}}(p)}{2 T}}{2 \omega_{\text{0}}(p)}
\widehat{\; 2k \;}^2 \widehat{k}^2 \widehat{p}^2,  \label{Vint}\\
\begin{array}{l}
\vspace{1.3cm} 
\includegraphics[scale=0.5]{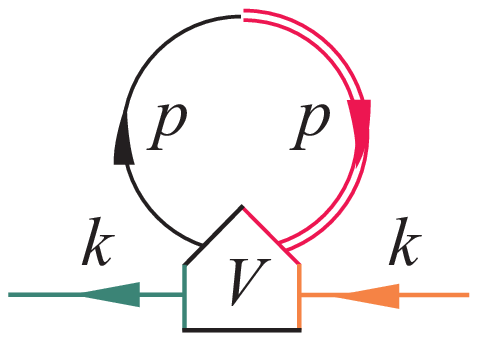} 
\end{array}  
& = &- 4 v_0 \int_p \frac{\coth\frac{\omega_{\text{0}}(p)}{2 T}}{2 \omega_{\text{0}}(p)}
\widehat{k}^2 \widehat{k}^2 \widehat{\; 2 p \;}^2. \nonumber
\eea
%-------------------------------------------------------------------------------------------------------------------
since each of diagrams listed above appears four times.

\subsubsection{KYA self-energy}

As a first step, we introduce the second-neighbor couplings
in (\ref{H0Lor}) and (\ref{HintLor}) by the
replacement $\nabla^2 \longrightarrow \nabla^2 - G^2  \nabla^4$ , where
%-------------------------------------------------------------------------------------------------------------------
\bea
 \nabla^4 := \nabla^2 \nabla^2  = \frac{4}{a^2} 
 \left[\nabla^2_{(2)} -\nabla^2 \right]  \label{Nabla_4},
\eea
%-------------------------------------------------------------------------------------------------------------------
$\nabla^2_{(2)}$ being defined in (\ref{LattLapl_1&2}) and the strength
of the next-nearest neighbor (NNN) coupling $G$ will be determined in perturbative
expansion. 
Much of the results obtained for the NN model still hold.
In particular, the propagator and the one-loop self energy
retain their general form given in (\ref{PropLor}) and 
the first line of (\ref{OneLoopSigmaNN}), respectively,
with free dispersion being
%-------------------------------------------------------------------------------------------------------------------
\bea
\Omega_0^2(k) = c^2 \left[ \widehat{k}^2 +  G^2 \widehat{k}^2  \widehat{k}^2 \right], 
\hspace{1cm} c^2 = \gamma^2/F^2. \label{Omega0G}
\eea
%-------------------------------------------------------------------------------------------------------------------

Further, let us now introduce additional   interaction terms
%-------------------------------------------------------------------------------------------------------------------
\bea
H_{\text{int}}^{(2)} & = & K^{(2)} c^2 v_0 \sum_x
(\bm \pi \cdot  \nabla^4 \bm \pi )
(\nabla^2 \bm \pi  \cdot \nabla^2 \bm \pi ) \nonumber  \\
H_{\text{int}}^{(3)} & = &  K^{(3)} c^2 v_0 \sum_x (\nabla^2 \bm \pi \cdot \nabla^2 \bm \pi)
(\nabla^2 \bm \pi  \cdot \nabla^2 \bm \pi) \nonumber \\
H_{\text{int}}^{(4)} & = &  A c^2 v_0 \sum_x  \bm \pi \cdot \nabla^2 \bm \pi \label{Hint123} \\
H_{\text{int}}^{(5)} & = & B c^2 v_0 \sum_x\bm \pi \cdot \nabla^4 \bm \pi \nonumber 
\eea
%-------------------------------------------------------------------------------------------------------------------
where  $H_{\text{int}}^{(3)}$ and $H_{\text{int}}^{(4)}$ serve as counter-terms, i.e. 
they absorb renormalizations
of $\nabla^2$ and $\nabla^4$ couplings.  

Individual contributions of (\ref{Hint123}) to the
self-energy are evaluated using colored diagrams, as outlined above.
To remain within KYA,   we discard a term proportional to 
$\widehat{k}^2 \widehat{k}^2 \widehat{k}^2$ and arrange the remaining
ones in two groups: the ones proportional to $\widehat{k}^2$ and
those proportional to $\widehat{k}^2 \widehat{k}^2$. The corresponding
self-consistent equation for magnon energy reduces to KYA form (\ref{KYAGF})
if $A=251/48$, $B+G^2 = 6 a^2 \widetilde{c}_0$, $K^{(2)} = a^6 \widetilde{W} /16$,
$K^{(3)} = -7 a^6 \widetilde{W} /64$, $F^2 G^2 = a^2 \widetilde{W} /8$
and $\gamma^2 = c^2 F^2$.
From the compact form of  magnon energies [equation (\ref{OmegaKJA})], 
one readily shows that $\widetilde{c}_1 = \widetilde{c}_2 = 1/2$ at $T=0$ K
so that the magnon dispersion becomes identical with standard LSW
result,  $\omega(k) = J(a^2/2) \widehat{k}^2$.
This is to be compared with
$\Omega_0 \propto  k$. Thus, in contrast to the perturbation
theories for type B magnons from preceding section,  interactions
substantially change the  magnon  dispersion law in the present case.
The possibility that interactions between Goldstone bosons
may have such a strong impact on the geometry of their dispersion
was anticipated only recently \cite{BraunerWatanabe}.
To the best of author's knowledge,   calculations
from this section offer the first explicit realization of  such a
mechanism. At finite temperatures,  we find
%-------------------------------------------------------------------------------------------------------------------
\bea
\widetilde{c}_1 &\approx& \frac 12 - \frac{3 I_{3/2}}{2}T^{\frac 32}-2 T^2, \;\;\;\;
\widetilde{c}_2 \approx \frac 12 - \frac{9 I_{3/2}}{2}T^{\frac 32}+2 T^2, \nonumber \\
I_{3/2}&=&\frac{\zeta(3/2)}{\sqrt{2 \pi}},  \label{TildeC}
\eea
%-------------------------------------------------------------------------------------------------------------------
leading to
%-------------------------------------------------------------------------------------------------------------------
\bea
\omega^2(k) \approx    4 J^2 a^2 T^2 k^2 + \frac{J^2 a^4}{4} \left[ 1 - 3 I_{3/2} T^{\frac 32}  \right]k^4
\label{OmegaKYAk2ik4}
\eea
%-------------------------------------------------------------------------------------------------------------------
in agreement with \cite{KYSOGF}.

\subsubsection{KYA Thermodynamics and Quantum Monte Carlo Simulation} \label{QMCA}

%-------------------------------------------------------------------------------------------------------------------
\begin{figure}
\bc 
\includegraphics[scale=0.75]{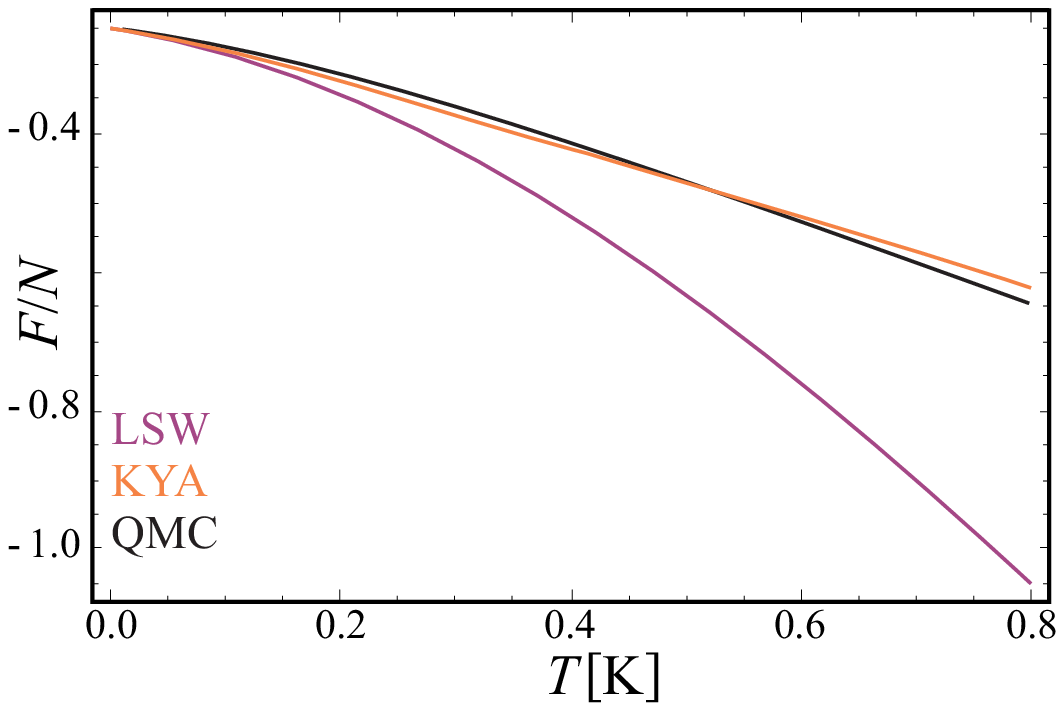}
\caption
{\label{fig4} The influence of  interactions on the free energy
of $S=1/2$ and $J = 1$ Heisenberg ferromagnet on the chain lattice.
The purple  curve represents the standard non-interacting model (LSW), 
while the black one denotes  QMC results. 
The orange  line corresponds to the model with interactions
of KYA type.}
\ec
\end{figure}
%-------------------------------------------------------------------------------------------------------------------

Now we discuss the  low-temperature series
for free energy in self-consistent KYA. According to (\ref{OmegaKYAk2ik4}) and
(\ref{EffLagrLor}), magnon fields admit standard type A Fourier decomposition including both creation
and annihilation operators which produces so-called zero-point term
in expression for free energy \cite{Kapusta}. 
The zero-point term is temperature dependent in the present case
 [see (\ref{OmegaKJA})] and,
since there are two  type A magnons, we find
%-------------------------------------------------------------------------------------------------------------------
\bea
\frac{F_{\text{KY}}^0}{N} & = & \frac{F_0}{N} + T v_0 \int_{p} 
\ln \left[ 1- \e^{-\omega_0(p)/T} \right]   \nonumber \\ 
&+& v_0 \int_p \omega_{\text{KY}}(p), \label{F0KYA}
\eea
%-------------------------------------------------------------------------------------------------------------------
where $F_0/N = -5/4$ provides a proper normalization at $T=0$K. 
To see that (\ref{KYAGF}) indeed reproduces all main features of 
low-temperature series for free energy in KYA, we 
calculate (\ref{F0KYA}) with the full Kondo-Yamaji dispersion \cite{KYSOGF}
and the
approximate one [equation (\ref{KYAGF}) and solutions (\ref{TildeC})]. The first
two terms of these series match completely. While the first one,
proportional to $T^{3/2}$,  is rather expected \footnote{The $T^{3/2}$ term is also found using
thermodynamic Bethe ansatz equations \cite{TakahashiBete},  
effective field theory \cite{Hofmann6},  Schwinger boson
mean field theory (SBMFT) \cite{Auerbach} and LSW}, the second one is proportional
to $T^2$ and  describes the effects of interactions. The 
$T^2$ term is twice the one obtained by thermodynamic Bethe ansatz equations \cite{TakahashiBete}
since KYA describes O(3) ferromagnet in terms of 
type-A magnons (a similar problem with $T^2$ term arises in SBMFT \cite{Auerbach}).
The direct application of TGF method yields $(7/3)T^2$ \cite{Koma} instead of $2 T^2$
and this discrepancy may be classified  as another artifact of TGF formalism.

To check the validity of KYA result for free energy  of a $S=1/2$ ferromagnetic chain
at arbitrary temperatures, we test it against QMC simulation. The simulation
was conducted using Wang-Landau algorithm and ALPS libraries \cite{ALPS_J_Stat_Mech}.
Results reliable up to $T\approx 0.01$K are obtained be  setting the cutoff at $3 \times 10^4$.

The results  shown at Fig. \ref{fig4} compare   KYA [equation (\ref{F0KYA})] to LSW  and QMC.
We see that, in contrast to naive LSW, KYA result agrees with QMC rather well in wide temperature range.
Further, the present exposition reveals the mechanism behind KYA: similarly to RPA, magnon-magnon
interactions generated by WZ term do not contribute to KYA. Instead, the dynamics of a
ferromagnetic chain is governed by NN and NNN couplings for type A magnons, combined
in a such manner that one-loop corrections yield spectrum usually encountered
at type B systems at $T=0$. Even though the KYA deviates from QMC only slightly,
which makes it a relatively good method for qualitative description
of a $S=1/2$ ferromagnet chain,
it introduces spurious terms arising from magnon-magnon interactions
in the low-temperature series for free energy already at leading order, i.e. at $T^2$
\cite{Hofmann7,TakahashiBete}.

%===============================================================================================================================================
\section{Sumarry}\label{Zaklj}
%===============================================================================================================================================

Linearization of equations of motion (EOM) for spin operators
is a popular and important tool for studying magnetic systems.
These include not only insulating materials with localized spins but also 
itinerant electron systems where Heisenberg Hamiltonian arises through a mapping on an
effective model. 
If the parameters of linearization are chosen carefully,
i.e. if an appropriate "decoupling scheme" is used, this method will provide reliable
results in many cases. 
For instance, solutions obtained with standard linearizations
 obey Mermin-Wagner theorem and agree with Monte Carlo simulations
and experimental results. 
However, in choosing  parameters of lineaarization one
must rely on physical intuition since approximations are not
controlled perturbativley (equations of motion contain
the Heisenberg-picture spin operators) nor do magnon operators
appear at any stage,  so all linearizations are generally considered to be \emph{ad hoc}.
This raises several questions concerning proper interpretation of results
obtained by EOM.
For example, it is hard to identify types of interactions generated
by various linearizations and their influence on thermodynamic
properties, or to separate interaction-induced effects from those
related to the choice
of dynamical degrees of freedom. Because complete understanding of EOM
method is a prerequisite for its successful application,
these issues need to be resolved.

On the other hand, effective field theory (EFT) is a powerful tool
for handling dynamics of Goldstone particles as it provides systematic (and
model independent) method for describing interaction effects.
From this point of view, O(3) ferromagnet is a rather interesting system
since magnon-magnon interactions come in two categories: the ones
induced by the unimodular constrain and those generated by the Wess-Zumino
(WZ) term. The main advantage of EFT over EOM is that magnon-magnon interaction
terms appear explicitly in the Hamiltonian (Lagrangian) making  the dynamics
of magnons  completely unrelated to the local su(2) algebra of spin
operators that define the Heisenberg Hamiltonian.

The present paper deals with interaction effects induced by the three
most commonly used linearizations: random phase approximation (RPA), 
Callen approximation (CA) and Kondo-Yamaji approximation (KYA). 
It is shown that all unique properties of linearizations mentioned above  
originate in different treatments of WZ term, where type B (A) magnons are
described by the model with (without)  WZ term. Since  all linearizations
discussed in the present paper respect discrete structure
of the Heisenberg Hamiltonian, we employ lattice regularization that
preserves these symmetries.
Perturbative calculations are conducted with the aid of colored diagrams which
are particularly suited for theories with derivative couplings.

As the simplest of type B models, RPA describes the O(3) ferromagnet
with one-loop interaction theory in such way that interactions specific
to the WZ term are neglected [see Sec. \ref{SecRPA}]. This simplification induces  violation of O(3) 
symmetry at leading interaction terms of order $\bm p^2$ and eventually produces
spurious terms in low-temperature series for free energy. 
Even though the the O(3) symmetry is explicitly broken by interaction
terms in effective Hamiltonian, the constrain 
$v_0 \int_{\bm p} \widehat{\bm p}^2 \Lan n_{\bm p} \Ran_0 =0$ in
the one-loop self energy
eliminates the gap at the same time introducing site independent
average number of magnons.
Thus, opposed to some claims found in the literature 
[see the discussion in the last paragraph of Sec. \ref{SecRPA}], magnon-magnon
interactions are accounted for  in RPA. These are interactions induced by the
unimodular constrain, i.e. they are generated by the 
geometry of coset space O(3)/O(2) = $S^2$.

The next approximation considered in the text is CA, viewed by many 
as an improvement over RPA. Unfortunately, the claims of superiority
of CA over RPA have been
supported only by some vague statements concerning additional 
correlations or interactions induced by this 
linearization scheme [see the discussion at the ending of Sec. \ref{LowTCA}], or by a subsequent
analysis of low-temperature series for spontaneous magnetization.
However, \emph{a posteriori} analysis of low-temperature series can not
distinguish between interaction effects and pure artifacts
of the formalism and therefore it can not be considered as a
complete one. In contrast, analysis based on the effective field theory
presented in Sec. \ref{SecCA} reveals that, regarding magnon-magnon
interactions, CA does contain certain improvements over RPA. Concretely,
four magnon interactions arising from WZ term appear in CA through
one-loop corrections in the magnon self-energy [see Sec. \ref{MagSpecCA}]. However, additional spurious
six and eight magnon interaction terms included in CA
violate O(3) symmetry at $\bm p^2$ through two and three loop corrections
to the self energy. Therefore,  the constrain 
$v_0 \int_{\bm p} \widehat{\bm p}^2 \Lan n_{\bm p} \Ran_0 =0$ must be 
imposed on those contributions to the  self energy that arise from  symmetry
violating terms, just as it is  in RPA. Also, one should note that all corrections
to the self energy in CA come from one-vertex diagrams.
Because of that, magnon energies in CA, as well as in RPA, posses
the same geometry as LSW solution (that is, $\omega(\bm k) \propto \widehat{\bm k}^2$).
Alternatively, as described in Sec. \ref{LowTCA}, magnon-magnon
interactions induced by WZ term and unimodular constrain
constitute two-loop corrections to the free energy in CA which
gives correct results at order $T^5$ (this is the leading
term in low-temperature series generated by interactions). However, 
symmetry violating terms produce deviation from true magnon interaction theory at
order $T^{11/2}$ (that is, at three-loop corrections to the free energy).

In addition, the applicability of lattice  field theories for type B magnons
in wide temperature range is demonstrated by comparison with quantum Monte Carlo (QMC)
simulation [see Sec. \ref{QMCB}]. Excellent agreement between  QMC and
LSW theory combined with one-loop correction to the free energy 
[equations (\ref{FLSW}) and (\ref{DeltaFTMIT})] additionally 
supports lattice EFT as a right method for interpretation of EOM results.

Further, we discuss KYA and show that its physics near $T=0$K can be reduced to
the self-consistent perturbation theory for type A magnons. Therefore, WZ term
and magnon-magnon interactions that it generates do not enter
KYA, making this approximation substainly different from RPA and CA.
Still, at $T=0$K, KYA reproduces standard  dispersion for
ferromagnetic magnons, $\omega(\bm k) \propto \widehat{\bm k}^2$.
This property of KYA is usually explained with phenomenological
vertex parameters which account spin-spin correlations
and maintain rotational invariance. In Sec. \ref{SecSOA} we show
that these  characterizations of KYA may be 
replaced  with  self-consistent one-loop perturbation theory for
a model of type A magnons with explicit identification
of interaction terms. In particular, KYA  Hamiltonian contains
lattice Laplacians that connect first as well as second neighbors 
on a chain lattice and these additional interactions play a key
role in producing characteristic dispersion at $T=0$K.
Analysis of the free energy reveals that KYA low-temperature series for
free energy deviates from rigorous results already at leading
term which is a consequence of interactions.
However, at the same time, KYA result differs from
QMC only slightly in wide temperature range [section \ref{QMCA}]
and provides qualitative description of the low-temperature thermodynamics.

A detailed analysis based on lattice magnon fields
presented in the paper revealed physics behind  RPA, CA and KYA
in case of O(3) ferromagnets. All three approximations are
reduced to equivalent systems of interacting magnons thus providing better
understanding of widely used theoretical tools.

\section*{Acknowledgement}

This work was supported by the Serbian Ministry of Education and Science
under Grant No. OI 171009.

\bibliography{Refs}

\end{document}